\documentstyle[12pt,epsf,axodraw]{article}
\newcommand{\la}[1]{\label{#1}}
\newcommand{\be}{\begin{equation}}
\newcommand{\ee}{\end{equation}}
\newcommand{\ba}{\begin{eqnarray}}
\newcommand{\ea}{\end{eqnarray}}

\newcommand{\nr}[1]{(\ref{#1})}
\newcommand{\msbar}{\overline{\mbox{\rm MS}}}
\newcommand{\Tint}[1]{{\hbox{$\sum$}\!\!\!\!\!\!\int}_{\!\!\!\!#1}}
\newcommand{\lsim}{\raise0.3ex\hbox{$<$\kern-0.75em\raise-1.1ex\hbox{$\sim$}}}
\newcommand{\gsim}{\raise0.3ex\hbox{$>$\kern-0.75em\raise-1.1ex\hbox{$\sim$}}}
\makeatletter \@addtoreset{equation}{section} \makeatother
\renewcommand{\theequation}{\arabic{section}.\arabic{equation}}
\begin{document}

\newcommand{\nn}{\nonumber}
\newcommand{\tr}{{\rm Tr\,}}
\newcommand{\fr}[2]{{\frac{#1}{#2}}}
\newcommand{\bmu}{\bar{\mu}}
\newcommand{\tb}{\tan\!\beta}
\newcommand{\tbmu}{\tan\!\beta(\bmu)}
\newcommand{\pf}{\frac{1}{16\pi^2}}
\newcommand{\hc}{{\rm H.c.}}
\newcommand{\haha}{H^{1\dagger}H^1}
\newcommand{\hbhb}{H^{2\dagger}H^2}
\newcommand{\hahb}{H^{1\dagger}\tilde{H}^2}
\newcommand{\hbha}{\tilde{H}^{2\dagger}H^1}
\newcommand{\had}{H^{1\dagger}}
\newcommand{\thad}{\tilde{H}^{1\dagger}}
\newcommand{\hbd}{H^{2\dagger}}
\newcommand{\thbd}{\tilde{H}^{2\dagger}}
\newcommand{\cTT}{\frac{\zeta(3)}{128\pi^4T^2}}
\newcommand{\saa}{\sin\!2\alpha\,}
\newcommand{\caa}{\cos\!2\alpha\,}
\newcommand{\ssa}{\sin^2\!\alpha\,}
\newcommand{\cca}{\cos^2\!\alpha\,}
\newcommand{\sbb}{\sin\!2\beta\,}
\newcommand{\cbb}{\cos\!2\beta\,}
\newcommand{\ssb}{\sin^2\!\beta\,}
\newcommand{\ccb}{\cos^2\!\beta\,}
\newcommand{\figysize}{16.0cm}
\newcommand{\figtopspace}{\vspace*{-1.5cm}}
\newcommand{\figbottomspace}{\vspace*{-5.0cm}}
\newcommand{\lmu}{\ln\frac{\bmu^2}{m_Z^2}}

\begin{titlepage}
\begin{flushright}
HD-THEP-96-13\\
hep-ph/9605283\\
September 9, 1996
\end{flushright}
\begin{centering}
\vfill

{\bf 
EFFECTIVE THEORIES OF MSSM AT HIGH TEMPERATURE}
\vspace{1cm}

M. Laine\footnote{m.laine@thphys.uni-heidelberg.de} \\

\vspace{1cm} {\em 
Institut f\"ur Theoretische Physik, 
Philosophenweg 16, 
D-69120 Heidelberg, Germany}

\vspace{2cm}

{\bf Abstract}

\vspace{0.5cm}
We construct effective 3d field theories for the Minimal 
Supersymmetric Standard Model, relevant for the thermodynamics of 
the cosmological electroweak phase transition. The effective theories
include a 3d theory for the bosonic sector of the original 4d theory; 
a 3d two Higgs doublet model; and a 3d SU(2)+Higgs model. The 
integrations are made at 1-loop level. In integrals related to 
vacuum renormalization we take into account only quarks and squarks 
of the third generation. Using existing non-perturbative lattice 
results for the 3d SU(2)+Higgs model, we then derive infrared safe 
upper bounds for the lightest Higgs boson mass required for successful 
baryogenesis at the electroweak scale. The Higgs mass bounds turn out 
to be close to those previously found with the effective potential, 
allowing baryogenesis if the right-handed stop mass parameter $m_U^2$ 
is small. Finally we discuss the effective theory relevant for $m_U^2$ 
very small, the most favourable case for baryogenesis.
\end{centering}

\vspace{0.3cm}\noindent

\vfill \vfill
\noindent

\end{titlepage}

\section{Introduction}
\la{intro}

The generation of the baryon number of the
Universe remains to be satisfactorily explained. It is quite 
plausible, though, that an important role in the process was 
played by the cosmological electroweak phase transition~\cite{krs}. 
Within the Standard Model the phase transition appears nevertheless 
to be too weakly of first order 
to produce the baryon asymmetry for realistic 
Higgs masses (for a review, see~\cite{rs}). 
Additional problems may be related to the 
amount of CP-violation available. Hence one is led to extensions 
of the Standard Model. 

One possible consistent extension of the Standard
Model is the Minimal Supersymmetric Standard Model (MSSM).
It has a large parameter space available so that it should 
be possible to find some corner with a strong enough first
order transition.
In addition, there are additional sources of CP-violation.
Indeed, the electroweak phase transition in MSSM has been
studied quite actively~[3--7]\footnote{
Upon completion of this work, three more papers on the 
same subject appeared~[37--39]. 
In~\cite{dggw}
the 1-loop effective potential is studied. In~\cite{ck,lo}
the authors study the dimensional reduction of MSSM as 
in the present work. 
In~\cite{ck} the analysis is a bit less complete than here 
and the conclusions are somewhat different. In~\cite{lo} the formulas
for dimensional reduction and heavy scale integrations are 
in some parts more, in some parts less complete than here, 
but vacuum renormalization and the 
implications of the formulas to the electroweak phase 
transition are not discussed.}. 

The investigations made so far (apart from~\cite{ck,lo}) 
have been based on the 
1- and 2-loop effective potentials for the Higgs field. 
The limit that the CP-odd Higgs 
mass $m_A$ is infinite was taken in~\cite{mgi,eqz,cqw,e},
leaving just one Higgs doublet and  
being the most favourable case for baryogenesis~\cite{beqz}.
The result of these investigations was that in general, 
it appears difficult to make a strong enough transition
unless the right-handed soft supersymmetry breaking 
stop mass parameter $m_U^2$ is small. Recently it has been 
noted that even smaller values of $m_U^2$ than originally
considered should be phenomenologically possible~\cite{cqw}, 
leading to a transition that is definitely strong
enough for baryogenesis. In addition, 
2-loop effects have been found to be favourable~\cite{e}. 

All the studies based on the effective potential
are subject to the
infrared (IR) problem at finite temperature~\cite{ir}. 
The IR problem
is related to the zero Matsubara components of bosonic 
fields, and precisely these components account
for the cubic 1-loop terms in the effective
potential studied in~[3--6], as well 
as for the logarithmic 2-loop terms making
the effect in~\cite{e}. The IR-problem calls 
for non-perturbative investigations of the problem. The 
method of choice for non-perturbative investigations is
the framework of dimensional reduction~[9--17].
Dimensional reduction means that one constructs
an effective 3d theory producing the same Green's
functions as the original
theory for the light bosonic fields. 
The perturbative dimensional reduction
step is free of IR-problems, and the resulting 
super-renormalizable 3d theory can 
then be studied with high precision Monte Carlo 
simulations~[18--22]. 

The non-perturbative investigations of the electroweak
phase transition in the Standard Model have
revealed the following pattern~\cite{klrs2}.
As long as the transition is strong enough for
baryogenesis, the IR-problems are not very dramatic
and effective potential studies do produce a reasonable
estimate of the properties of the phase transition. 
When the transition gets weaker, non-perturbative
effects become large. However, even if non-perturbative
effects are small for stronger transitions, 
it is nevertheless interesting to note 
that prior to the lattice study in~\cite{klrs2}
many perturbative studies stated that baryogenesis
was possible up to $m_H\sim 45$ GeV. In~\cite{klrs2}
it was discovered that practically {\em no Higgs mass}
is possible. While this effect is mostly related to 
vacuum renormalization instead of non-perturbative IR-effects, 
it nevertheless proves that it is important to work 
in a consistent framework where all the approximations 
are under control.

The purpose of the present paper is to make
a dimensional reduction for the MSSM. We also 
perform further integrations inside the dimensionally
reduced 3d theory, to arrive at the simplest possible
effective theory. In particular, we construct a 3d
two Higgs doublet model and a 3d SU(2)+Higgs model
in the part of the parameter space where it is possible.
For the latter theory, the existing
non-perturbative lattice results  
allow to remove the IR problem from the Higgs mass bound.
The bound derived is in principle also gauge and 
$\bmu$-independent, unlike the ratio $v(T_c)/T_c$ derived from 
the effective potential.

On the technical side, one purpose of the present investigation
is to study how the cubic scalar vertices, not present 
in the Standard Model, affect dimensional reduction.

In comparison with~[4--7], we also try 
to be more explicit about the effects of vacuum 
renormalization. The theory studied is more or less the same. 
We include here the bottom Yukawa coupling $h_b$
and study a general CP-odd Higgs mass $m_A$ as in~\cite{beqz}.

It is found that in the region of the parameter
space where reduction into the 3d SU(2)+Higgs model
is possible and the transition is strong enough 
for baryogenesis, the non-perturbative results 
agree with the effective potential
investigations. The conclusion is that Higgs masses  
$m_h\lsim 75$ GeV produce a strong enough
transition if $m_U^2$ is small enough, $m_U^2\lsim (50-100)^2$ GeV$^2$.
Hence, the situation has improved with respect to 
the Standard Model where no Higgs mass is possible. 
Where reduction into SU(2)+Higgs cannot 
be made --- notably when
$m_U^2$ is still smaller and the transition is even
stronger~\cite{cqw,e} --- we propose an effective 3d theory 
allowing more detailed studies of the problem.

The plan of the paper is the following.
In Sec.~\ref{sm} we briefly review the Higgs mass
bound in the Standard Model and its derivation within the 3d framework.
In Sec.~\ref{lagrangian} we state in some detail the approximations
adopted and the Lagrangian used in the present investigation. 
Sec.~\ref{dimred} contains the dimensional reduction into a 3d
bosonic effective theory. In Sec.~\ref{squarks} we make further 
integrations inside the 3d theory, removing the squarks and the temporal 
components of the gauge fields. The resulting two Higgs doublet
model is diagonalized in Sec.~\ref{diag}, and the heavy Higgs doublet
in integrated out in Sec.~\ref{heavyH}. In Sec.~\ref{vacren} we 
discuss how the running Lagrangian parameters are fixed 
through vacuum renormalization. The numerical results for
the strength of the transition are in Sec.~\ref{numres}.
Finally, in Sec.~\ref{Utheory} we propose an effective
theory for describing the phase transition if the mass parameter
$m_U^2$ is very small. Sec.~\ref{conclu} is the conclusions.

\section{The EW phase transition in the Standard Model}
\la{sm}

The thermodynamics of the
electroweak phase transition in the full Standard Model
has been extensively studied in the literature (\cite{rs}
and references therein). Perturbative studies exist up to 2-loop 
level~\cite{ae,fh,fkrs1,kls1}. Non-perturbative lattice studies 
rely on perturbative 2-loop dimensional 
reduction~[12--15], 
and have been performed for a wide
range of Higgs masses~[18--22]. 
The Higgs
mass bound in terms of the parameters of the 3d SU(2)+Higgs
model was derived in~\cite{klrs2}.

The Higgs mass bound arises as follows. 
Assume that there in some underlying physical 4d theory
in which the electroweak phase transition takes place so 
that the static Green's functions of the lightest
excitations are described by the effective theory\footnote{
It should be noted that even though only the SU(2) group
is displayed explicitly in eq.~\nr{L3d}, the perturbative
effects of the U(1) group, making the phase transition 
stronger, have been included in the bound~\nr{bound}. No non-perturbative
lattice simulations exist yet for the SU(2)$\times$U(1)+Higgs
theory.}
\ba
L_{\rm 3d} & = & \fr14 F^a_{ij}F^a_{ij} 
+ (D_i\phi)^\dagger(D_i\phi)+
m_3^2 \phi^{\dagger}\phi+
\lambda_3 (\phi^\dagger\phi)^2, \la{L3d}
\ea
where  $D_i=\partial_i-ig_3\tau^a A^a_i/2$. 
Then the phase transition is strong 
enough for baryogenesis 
if at the phase transition point~\cite{klrs2}
\be
x\equiv\frac{\lambda_3}{g_3^2} < 0.03 - 0.04. \la{bound}
\ee
Since the theory in eq.~\nr{L3d} is super-renormalizable, 
the parameters $\lambda_3, g_3^2$ do not run and the quantity
$x$ is a well-defined pure number. It is also gauge-independent. 
The uncertainty in~\nr{bound}
arises from uncertainties in estimates of the sphaleron rate
in the broken phase and from uncertainties in the real-time 
dynamics of the phase transition (whether the Universe reheats
back to $T_c$ after the nucleation period, etc.).

For the Standard Model, the parameters $m_3^2$, $\lambda_3$ and
$g_3^2$ have been calculated in terms of temperature and the physical 
zero-temperature parameters of the theory in~\cite{klrs1}. Then one 
may solve for the critical temperature from the condition 
\be
m_3^2(m_h,T_c)=0, \la{Tc}
\ee
and use this $T_c$ in the estimate of 
\be
x=\frac{\lambda_3(m_h,T_c)}{g_3^2(m_h,T_c)}.
\ee
Eq.~\nr{Tc} does not give $T_c$ exactly (it corresponds
to resummed 1-loop accuracy), 
but this does not matter since $\lambda_3/g_3^2$ depends
on $T_c$ only through logarithmic 1-loop corrections. From an 
analysis of the type outlined, one gets that 
no Higgs mass (or at most an 
extremely light Higgs mass, $m_h\lsim 20$ GeV)
would satisfy the bound~\nr{bound} 
in the Standard Model since $x>0.04$ 
due to top Yukawa coupling corrections, 
see Fig.~27 in~\cite{klrs2}.

In this paper we study whether a theory of the 
type in eq.~\nr{L3d} can be constructed in the MSSM
and what would be the Higgs mass bound implied.

\section{The Lagrangian}
\la{lagrangian}

We start by discussing 
the Lagrangian used and the simplifications made, fixing
at the same time the notation. We work throughout in Euclidian space
and for definiteness in the Landau gauge. The value of $x$
derived is gauge-independent.

The main simplifications are the following 
(for the complete Lagrangian in MSSM, see~\cite{r}). 
First, we neglect  
the U(1) subgroup in loop corrections related to 
vacuum renormalization and dimensional reduction. 
That is, no difference is made between $g^2$ and $g^2+g'^2$ 
beyond tree-level in IR-safe integrals. 
This is a good approximation as far as the electroweak phase transition 
is concerned, especially with respect to the other 
uncertainties in the calculation. Even at tree-level, 
we display explicitly only the covariant 
derivatives related to SU(2) and SU(3). 

Second, we will assume that the gaugino and 
higgsino mass parameters in the symmetric phase 
are so large that these fields
have decoupled, as is usually assumed in the present
context~[4--7]. 
Even if the masses are smaller, 
these fields do not have very much significance, being
fermions: at finite temperature, the important effects
arise from IR-sensitive bosons. In the framework of
the present paper, the extra fermions would only affect the
parameters of the 3d theory in the
first dimensional reduction step, but the later integrations
remain precisely the same. It should also be noted that 
gauginos and higgsinos do not couple to the scalar Higgs degrees
of freedom through the dominant Yukawa coupling $h_t$, 
unlike the top quark. In general, it is expected that 
the effect of gauginos and higgsinos would be to make 
the phase transition weaker, due to the increased screening
in the thermal masses~\cite{eqz,beqz,e}. However, 
gauginos and higgsinos do have an effect when 1-loop
corrections to the top Yukawa coupling $h_t$ are calculated. 
Since $h_t$ gives
the most important effects in the present calculation, 
the loop corrections may also be important. 
We return to this point in more detail below.

Third, only the squark partners of top and bottom
quarks are assumed to be light enough to affect the 
electroweak phase transition.

Then the remaining fields
are as follows: there are the SU(2) 
and SU(3) gauge fields $A^a_\mu$, $C^A_\mu$.
The Higgs fields are $H^1, H^2$,  with hypercharges $Y=-1,+1$.
The adjoint Higgs fields with opposite hypercharges are denoted by
\be
\tilde{H}^1=i\tau_2H^{1*},\quad 
\tilde{H}^2=i\tau_2H^{2*}.
\ee
The 2-index antisymmetric tensor is defined through
$\epsilon_{12}=-1$, so that 
$\tilde{H}^n_i=-\epsilon_{ij}H^{n*}_j$.
We use the notation
\be
H^1=
\left( \begin{array}{l} 
H^1_1 \\
H^1_2
\end{array} \right)=
\frac{1}{\sqrt{2}}
\left( \begin{array}{r} 
h^1_0 + i h^1_3 \\
-h^1_2 + i h^1_1
\end{array} \right), \quad
H^2=
\left( \begin{array}{l} 
H^2_1 \\
H^2_2
\end{array} \right)=
\frac{1}{\sqrt{2}}
\left( \begin{array}{l} 
h^2_2 + i h^2_1  \\
h^2_0 - i h^2_3
\end{array} \right) \la{h1h2}
\ee
for the complex and real components of the Higgs fields, 
so that at zero temperature
\be
\langle h^1_0\rangle=v_1,
\quad
\langle h^2_0\rangle=v_2. \la{broken}
\ee

The fermions of the third generation are 
\be
q_{L\alpha} = \left(
\begin{array}{l}
t_{L\alpha} \\
b_{L\alpha} 
\end{array}\right), \quad
t_{R\alpha},\quad b_{R\alpha},
\ee
where $\alpha$ is the SU(3)-index and the 
hypercharges are $Y=1/3, 4/3, -2/3$, respectively.
Correspondingly, the 
squarks of the third generation are
\be
Q_\alpha= 
\left( \begin{array}{l} 
\tilde{t}_{L\alpha} \\
\tilde{b}_{L\alpha}
\end{array} \right), \quad
\quad U_\alpha=\tilde{t}_{R\alpha}^*,
\quad D_\alpha=\tilde{b}_{R\alpha}^*, 
\ee
with the hypercharges $Y=1/3, -4/3, 2/3$.
The fields $U, D$ transform under SU(3) with the 
adjoint generators $\overline{\lambda}_A=-\lambda_A^*$.

The part of the action containing the kinetic terms of and 
interactions between gauge fields and fermions remains as in 
the Standard Model. For the Higgs and squark fields 
the quadratic terms are  
\ba
L & = & 
(D^w_\mu H^1)^\dagger(D^w_\mu H^1)+
(D^w_\mu H^2)^\dagger(D^w_\mu H^2) \nn \\
& + & m_1^2 H^{1\dagger}H^1+
m_2^2 H^{2\dagger}H^2+m_{12}^2 (H^{1\dagger}\tilde{H}^2+
\tilde{H}^{2\dagger}H^1) \nn \\
& + & 
(D^{ws}_\mu Q)^\dagger(D^{ws}_\mu Q)+
(D^s_\mu U^*)^\dagger(D^s_\mu U^*)+
(D^s_\mu D^*)^\dagger(D^s_\mu D^*) \nn \\
& + & 
m_{Q}^2 Q_\alpha^\dagger Q_\alpha+
m_{U}^2 U_\alpha^* U_\alpha+
m_{D}^2 D_\alpha^* D_\alpha,
\ea
where 
$D_\mu^{ws}=\partial_\mu-ig\tau^a A^a_\mu/2-
i g_S\lambda^AC^A_\mu /2$ and
$w$ and $s$ indicate the charge included.

The supersymmetric interactions are generated by 
the superpotential and by the $D$-terms. We take
the superpotential to be 
\be
W=\mu\epsilon_{ij}H^1_iH^2_j+h_t\epsilon_{ij}H^2_iQ_{j\alpha}U_\alpha
+h_b\epsilon_{ij}H^1_iQ_{j\alpha}D_\alpha.
\ee
Hence also the bottom Yukawa coupling $h_b$ is kept, 
although its effect is small since the region of parameter 
space which can affect baryogenesis is around $\tb=v_2/v_1\sim 2$, 
and $h_b\sim m_b/(2 m_Z \cos\beta)$.
The interaction Lagrangian following from the
superpotential is
\ba
L_{W} & =  &
h_t\Bigl(\bar{t}_R\tilde{H}^{2\dagger}q_L+
\bar{q}_L\tilde{H}^2t_R\Bigl)+
h_b\Bigl(\bar{b}_R\tilde{H}^{1\dagger}q_L+
\bar{q}_L\tilde{H}^1b_R\Bigl) \nn \\
& + & 
h_t^2\Bigl(
Q^*_{i\alpha}U^*_{\alpha}Q_{i\beta}U_\beta+
H^{2\dagger}H^2U^*_{\alpha}U_{\alpha}+
\tilde{H}^{2\dagger}Q_{\alpha}Q^\dagger_\alpha
\tilde{H}^2\Bigr) \nn \\
& + & 
h_b^2\Bigl(
Q^*_{i\alpha}D^*_{\alpha}Q_{i\beta}D_\beta+
H^{1\dagger}H^1D^*_{\alpha}D_{\alpha}+
\tilde{H}^{1\dagger}Q_{\alpha}Q^\dagger_\alpha
\tilde{H}^1\Bigr) \nn \\
& + & 
h_t h_b \Bigl(
H^{2\dagger}H^1U^*_{\alpha}D_{\alpha}+
H^{1\dagger}H^2U_{\alpha}D^*_{\alpha}
\Bigr) .
\ea

The interaction Lagrangian following from the $D$-terms,  
on the other hand, is
\ba
L_{D} & = & 
\frac{g'^2}{8}\Bigl[\haha-\hbhb\Bigr]^2 \nn \\
& + & 
\frac{g^2}{8}
\Bigl[Q^*_{i\alpha}Q_{j\alpha}+H^{m*}_iH^m_j\Bigr]
\Bigl[Q^*_{k\beta}Q_{l\beta}+H^{n*}_kH^n_l\Bigr]
\tau^a_{ij}\tau^a_{kl} \nn \\
& + & 
\frac{g_S^2}{8}
\Bigl[Q^*_{i\alpha}Q_{i\beta}-
U_\alpha U^*_\beta-D_\alpha D^*_\beta\Bigr]
\Bigl[Q^*_{j\gamma}Q_{j\delta}-
U_\gamma U^*_\delta-D_\gamma D^*_\delta\Bigr]
\lambda^A_{\alpha\beta}
\lambda^A_{\gamma\delta}. \la{Dterm} 
\ea
Here we kept the U(1) coupling $g'$ only in the Higgs sector.
In eq.~\nr{Dterm}, 
\be
\tau^a_{ij}\tau^a_{kl}=(2\delta_{il}\delta_{jk}-
\delta_{ij}\delta_{kl}),\quad
\lambda^A_{\alpha\beta}
\lambda^A_{\gamma\delta}=
\fr23
(3\delta_{\alpha\delta}\delta_{\beta\gamma}-
\delta_{\alpha\beta}\delta_{\gamma\delta}).
\ee

The soft supersymmetry breaking cubic
interactions are
\be
L =
u_{s}\thbd Q_\alpha U_\alpha+
d_{s}\thad Q_\alpha D_\alpha+
e_{s}\hbd Q_\alpha D_\alpha+
w_{s}\had Q_\alpha U_\alpha+\hc
\ee 
Here we use the notation 
of~\cite{r}, with opposite signs. 
The relation to the more standard $A$-parameters is discussed 
in Sec.~\ref{vacren} in connection with vacuum renormalization.
In particular, it should be noted that the parameter
$e_s$ includes the term $\mu h_b$ arising from 
the superpotential and the parameter $w_s$ includes
$-\mu h_t$. Since there is also an arbitrary soft
component in these terms~\cite{r} and since we 
assume the higgsinos to be so heavy that they have 
decoupled, the theory does in fact not depend at all 
on the true supersymmetric mass parameter $\mu$.
In the following, we shall restrict the 
parameters $u_s, d_s, e_s, w_s$ to be real.

Although quite a few simplifications have been made,
there are still a lot more parameters left than in the Standard
Model. The two scalar sector parameters
$\nu^2, \lambda$ appearing there are replaced by
$m_1^2, m_2^2, m_{12}^2, m_Q^2, m_U^2, m_D^2, u_s, d_s, e_s, w_s$.

There is the following important point to be noticed
about the coupling constants $g, g_S, h_t, h_b$ in 
the present theory. As an example, take the gauge coupling.
If one takes the theory under investigation as such, 
then the weak gauge coupling in the gauge sector runs as
\be
\bmu\frac{d g^2(\bmu)}{d\bmu}=
\frac{g^4}{8\pi^2}\frac{8 n_F+N_s-44}{6},
\ee
where $n_F=3$ is the number of families and 
$N_s=5$ is the number of scalar doublets interacting 
with the SU(2) gauge fields. However, the couplings 
$g_1^2, g_2^2, g_3^2, g_4^2=g^2$ in the 
SU(2)-part of the scalar potential
following from eq.~\nr{Dterm}, 
\be
V =
\fr18 g_1^2({H^1}^\dagger H^1)^2+
\fr18 g_2^2({H^2}^\dagger H^2)^2+
\fr14 g_3^2 {H^1}^\dagger H^1 {H^2}^\dagger H^2 -
\fr12 g_4^2 {H^1}^\dagger \tilde{H}^2 \tilde{H}^{2\dagger} H^1,
\ee
run as 
\ba
\bmu\frac{d g_1^2(\bmu)}{d\bmu} & = & 
\bmu\frac{d g_2^2(\bmu)}{d\bmu} = 
\bmu\frac{d g_3^2(\bmu)}{d\bmu} = 
\frac{7}{2}\frac{g^4}{8\pi^2}, \\
\bmu\frac{d g_4^2(\bmu)}{d\bmu} & = &  
-\frac{5}{2}\frac{g^4}{8\pi^2}.
\ea
Hence within the present theory 
one would have to renormalize these couplings 
separately from the coupling in the gauge sector.
In other words, one has to consider
a large number of zero-temperature observables 
in terms of which to fix the independent parameters.
If on the other hand one
wants to maintain the universality of the gauge coupling, 
then one has to include 
the complete supersymmetric structure of the 
theory in the calculation in one way or the other. 
In the present theory, supersymmetry
is maintained only in the quark-squark sector of the third 
generation, and indeed, if only these fields are included in 
the internal lines of loop integrals, then $g^2$ runs 
everywhere as
\be
\bmu\frac{d g^2(\bmu)}{d\bmu}  =  
\frac{3}{2}\frac{g^4}{8\pi^2}
\ee
We shall work within the accuracy of this approximation here and 
assume the gauge coupling to be universal.

The same thing applies also to the Yukawa couplings $h_t, h_b$ 
and is quite important there as well, since $h_t$ is large and gives
the dominant effects. Indeed, within the present theory, the 
Yukawa couplings in the different
squark-Higgs and quark-Higgs interactions
run differently. To get a universal Yukawa coupling, higgsinos 
and gauginos should be included. However, we
will be satisfied with the present approximation in this paper 
for two reasons.
First, the Yukawa coupling $h_t$ is determined by the top mass 
which is not known very precisely at the moment. Second and even more 
important, the most significant effects of $h_t$ appear in 
conjunction with the soft squark mass parameters 
$m_Q^2, m_U^2, m_D^2$ (see below). Since these are unknown,   
there is a large uncertainty in the calculation in any case.
Once the squark masses have been measured and the top mass
is known more precisely, the gauge coupling $h_t(\bmu)$ should
be fixed at 1-loop level in terms of the top pole mass.

We will work in the $\msbar$ scheme (with the scale parameter $\bmu$) 
and take $\tr 1=4$. For the squark and quark 
loops included in vacuum renormalization
the results agree with those in the
$\overline{\mbox{\rm DR}}$-scheme,
often used in supersymmetric theories.

\section{Dimensional reduction}
\la{dimred}

Let us first recall the expansion parameters
of dimensional reduction~\cite{klrs1}. Since all the calculations 
are IR-safe, no non-analytic powers of masses can appear.
In fact, the expansion proceeds just in powers of
\be
\frac{h_t^2}{16\pi^2},\quad
\frac{g_S^2}{16\pi^2}
\ee
as at zero temperature.
We include only the quarks and squarks of the third generation
in the loops affecting vacuum renormalization, whereas
the corrections e.g.\ from gauge bosons,
suppressed by $g^2/(16\pi^2)$, are neglected.

In addition to expanding in coupling constants, we
make a high-temperature expansion in the mass parameters.
This requires that the soft supersymmetry
breaking mass parameters satisfy 
\be
m_1^2, m_2^2, m_{12}^2, m_U^2, 
m_D^2, m_Q^2 <
(2 \pi T_c)^2. \la{limit}
\ee
The limit~\nr{limit} implies that 
$m_A\lsim 2\pi T_c$, so that the results of this paper
cannot be directly continued to the limit $m_A\to\infty$
studied in~\cite{eqz,cqw,e}. Actually, the limitation
on $m_A$ is not as important as that on the squark mass
parameters, since the latter are associated with larger
coupling constants.

To keep track of the validity of the high-temperature
expansion, we will at some points display also
the leading correction terms. The
critical temperature is $T_c\sim 100$ GeV so that we
shall assume $m_Q, m_U, m_D, m_A \lsim 300$ GeV. We also
assume that the masses generated at the electroweak 
phase transition as well as the masses associated with
possible colour and charge breaking minima
are small compared with $2\pi T_c$. 

If some of 
the soft masses are large, one cannot use the high-temperature
expansion. Instead, one should
evaluate the corresponding integrals numerically. 
If $m\gg 2\pi T_c$, one can also 
integrate out these degrees of freedom in the
sense of the vacuum decoupling theorem~\cite{ac}.

The actual dimensional reduction proceeds by writing down
the general form of the effective 3d theory and then
determining the 3d coupling constants by
matching the Green's functions
in the original theory and in the 3d theory. The degrees of freedom 
of the effective theory are the bosonic degrees of freedom 
of the original theory. The temporal components
of gauge fields become Higgs fields in the adjoint representation.
The structure of the 3d theory is determined by 
gauge invariance. The 1-loop calculations needed 
are a straightforward application of the rules in~\cite{klrs1}.
We just write down the graphs and the results below.

Since the complete bosonic sector of MSSM is 
rather large, we display only the part interacting
with the SU(2) and Higgs degrees of freedom explicitly. We recall
that after trivial rescaling with $T$, the dimension
of bosonic fields in 3d is GeV$^{1/2}$ and that of
the couplings $g_3^2, g_3'^2, h_{t3}^2, h_{b3}^2$ is GeV.
At some points, we denote new parameters with the same
symbols as the old ones, to avoid increasingly 
cumbersome notation. Higher-order operators suppressed 
by the temperature and coupling constants
are neglected. Finally, 
let us recall some basic notation:
\ba
c_B & = & \ln(4\pi)-\gamma_E\approx 1.953808,\quad
c_F = c_B -2\ln 2\approx 0.567514 , \\
L_b(\bmu) & = & 2 \ln\frac{\bmu}{T}-2 c_B,\quad
L_f(\bmu) = 2 \ln\frac{\bmu}{T}-2 c_F. \la{Lb}
\ea

The 3d effective theory consists of the following parts:

1. The temporal components of the original gauge fields become scalar
fields in the adjoint representation, and the spatial components 
remain gauge fields. Relevant for the present discussion
is the part
\be
L_{\rm gauge}=
\fr14 F^a_{ij}F^a_{ij}+
\fr12 (D_iA_0^a)^2+\fr12 m_{A_0}^2 A_0^aA_0^a +
\fr12 (\partial_i C_0^A)^2 + \fr12 m_{C_0}^2 C_0^AC_0^A, 
\ee
where $D_iA_0^a=\partial_iA_0^a+g_3\epsilon^{abc}A^b_iA^c_0$.
When only quarks and squarks are included, the 
3d fields are related to the renormalized 4d fields by
\ba
(A_0^aA_0^b)^{({\rm new})} & = & \frac{1}{T}
(A_0^aA_0^b)(\bmu)\biggl[
1+\frac{g^2}{16\pi^2}\biggl(
[L_f(\bmu)-1]+\fr12[L_b(\bmu)+2]
\biggr)
\biggr], \\
(A_i^aA_j^b)^{({\rm new})} & = & \frac{1}{T}
(A_i^aA_j^b)(\bmu)\biggl[
1+\frac{g^2}{16\pi^2}\biggl(
L_f(\bmu)+\fr12 L_b(\bmu)
\biggr)
\biggr],
\ea
where mass corrections suppressed by $m_Q^2/(2\pi T_c)^2$ 
were neglected.

The gauge coupling can be most easily obtained from the graphs
(qqqq), (SS), (SSS1), (SSS2), (SSSS) in Fig.~\ref{fig:dimred}.c.
Here also a redefinition of the Higgs fields, given in
\nr{haha}--\nr{hbhb}, is needed.
After the redefinition one gets 
\be
g_3^2 = g^2(\bmu) T
\biggl[
1-\frac{g^2}{16\pi^2}\biggl(
L_f(\bmu)+\fr12 L_b(\bmu)
\biggr)
\biggr],
\ee
so that $g_3^2(A_i^aA_j^b)^{({\rm new})}=g^2 A^a_iA^b_j$.

The values of $m_{A_0}^2,m_{C_0}^2$ are well known~\cite{beqz,e}, 
but for completeness we write them here as well. These terms contain
only screening parts not related to vacuum renormalization, so that
we include the complete spectrum of the model in the loops.
With the notation in Fig.~\ref{fig:dimred}.a, the graphs 
contributing to $m_{A_0}$
are (ff), (gg), (AA), (HH), (QQ), (A), (H), (Q); 
to $m_{C_0}$ contribute (ff), (gg), (CC), (SS), (C), (S).
For illustration, the leading mass terms are also shown:
\ba
m_{A_0}^2 & =  & g^2
\biggl[
\biggl(\fr23+\frac{n_F=3}{3}+\frac{N_s=5}{6}
\biggr)T^2+\frac{1}{8\pi^2}
(m_1^2+m_2^2+3 m_Q^2)
\biggr], \la{ma0}\\
m_{C_0}^2 & =  & g_S^2
\biggl[
\biggl(1+\frac{n_F=3}{3}+\frac{N_s=4}{6}
\biggr)T^2+\frac{1}{8\pi^2}
(m_U^2+m_D^2+2 m_Q^2)
\biggr]. \la{mc0}
\ea 
In eqs.~\nr{ma0}, \nr{mc0}, $n_F=3$ is the number of fermion
families and $N_s$ is the number of scalar doublets interacting
with the gauge fields in question.

\begin{figure}[tb]

\figtopspace

\epsfysize=13.0cm
\centerline{\epsffile[100 320 500 750]{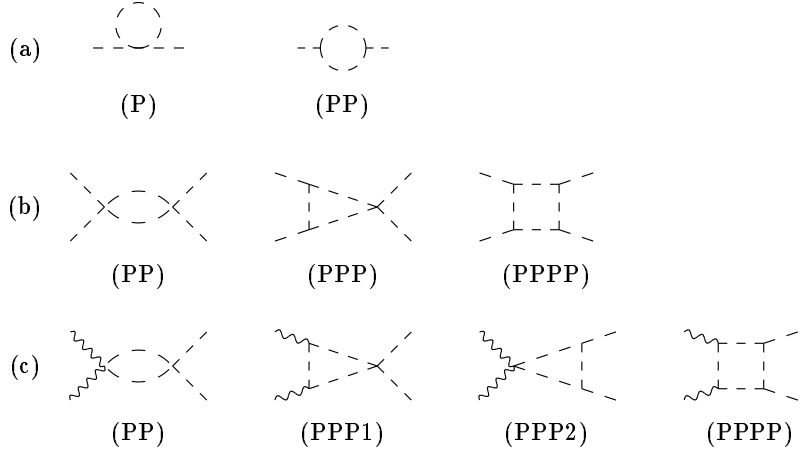}}

\figbottomspace

\caption[a]{
The generic types of graphs needed for dimensional reduction
of (a) wave functions and 
masses, (b) scalar couplings and (c) the gauge coupling.
Wiggly lines are vector propagators and dashed lines represent  
generic propagators of particle type P$=$Q, U, D, S, H, A, C, g, q, f. 
Here Q, U, D denote the corresponding
squarks, S is a squark in general, H is a Higgs doublet, A and C 
are the SU(2) and SU(3) gauge fields, g is a ghost, q a third
generation quark and f a general fermion. For the coupling 
constants, 1-loop dimensional reduction is directly
related to 1-loop vacuum renormalization and hence 
only squarks and quarks are considered in the internal lines. 
For the masses, the thermal screening terms
proportional to $T^2$ are not related to vacuum renormalization
and hence we include all the modes with $m \lsim 2\pi T$
in the loops.}
\la{fig:dimred}
\end{figure}

2. The quadratic terms of the Higgs sector are
\ba
L_{\rm Higgs} & = & 
(D_iH^1)^\dagger(D_iH^1)+
(D_iH^2)^\dagger(D_iH^2)+
\xi \Bigl[
(D_iH^1)^\dagger(D_i\tilde{H}^2)+\hc
\Bigr] \nn \\
& + & m_1^2 H^{1\dagger}H^1+
m_2^2 H^{2\dagger}H^2+m_{12}^2 (H^{1\dagger}\tilde{H}^2+
\tilde{H}^{2\dagger}H^1), \la{LHiggs}
\ea
where $D_i=\partial_i-ig_3 \tau^a A^a_i/2$ and $g_3$ is 
the 3d gauge coupling.
The graphs contributing to the 2-point Higgs correlators are
(qq), (AH), (SS), (A), (H), (S) in Fig.~\ref{fig:dimred}.a.
According to the general strategy of including 
only quark and squark loops in terms related to 
vacuum renormalization, we 
neglect terms multiplied by $g^2$ everywhere
except for the screening parts proportional to $T^2$.  
The new fields are then related to the renormalized
4d fields in the $\msbar$ scheme by
\ba
(\haha)^{({\rm new})} & = & \frac{1}{T}
(\haha)(\bmu)\biggl[
1+\pf 3h_b^2 L_f(\bmu)+\frac{\zeta(3)}{128\pi^4}
\frac{d_s^2+w_s^2}{T^2}
\biggr], \la{haha} \\
(\hbhb)^{({\rm new})} & = & \frac{1}{T}
(\hbhb)(\bmu)\biggl[
1+\pf 3h_t^2 L_f(\bmu)+\frac{\zeta(3)}{128\pi^4}
\frac{u_s^2+e_s^2}{T^2}
\biggr]. \la{hbhb}
\ea
The mass parameters, on the other hand, are
\ba
{m_1^2}^{({\rm new})} & = &
m_1^2(\bmu)-\frac{3}{16\pi^2}
\Bigl[
h_b^2(m_Q^2+m_D^2)L_b(\bmu)+h_b^2 m_1^2 L_f(\bmu)+
(d_s^2+w_s^2)L_b(\bmu) 
\Bigr] \nn \\
& + & \biggl(
\fr34 h_b^2+\fr14 g^2
\biggr)T^2 \nn \\
& + & 
\frac{\zeta(3)}{128\pi^4T^2}
\Bigl[3 d_s^2(m_Q^2+m_D^2)+3 w_s^2(m_Q^2+m_U^2)
-(d_s^2+w_s^2)m_1^2
\Bigr], \la{mm13} \\ 
{m_2^2}^{({\rm new})} & = & 
m_2^2(\bmu)-\frac{3}{16\pi^2}
\Bigl[
h_t^2(m_Q^2+m_U^2)L_b(\bmu)+h_t^2 m_2^2 L_f(\bmu)+
(u_s^2+e_s^2)L_b(\bmu) 
\Bigr] \nn  \\
& + & \biggl(
\fr34 h_t^2+\fr14 g^2
\biggr)T^2 \nn \\
& + & 
\frac{\zeta(3)}{128\pi^4T^2}
\Bigl[3 u_s^2(m_Q^2+m_U^2)+3 e_s^2(m_Q^2+m_D^2)
-(u_s^2+e_s^2)m_2^2
\Bigr], \la{mm23} \\ 
{m_{12}^2}^{({\rm new})} & = & 
m_{12}^2(\bmu)-\frac{3}{16\pi^2}
\Bigl[
\fr12 (h_t^2+h_b^2) m_{12}^2 L_f(\bmu)+
(u_s w_s-d_s e_s)L_b(\bmu) 
\Bigr] \nn \\
& + & 
\frac{\zeta(3)}{128\pi^4T^2}
\Bigl[3 u_sw_s(m_Q^2+m_U^2)-3 d_se_s(m_Q^2+m_D^2) \nn \\
& - & \fr12 (u_s^2+d_s^2+e_s^2+w_s^2)m_{12}^2
\Bigr],  \la{mm123}
\ea
where we have shown terms up to quadratic order
in the masses. From the high-temperature expansion, 
one would also get terms of the form $h_t^2 m^4$
in addition to the $u_s^2 m^2$-terms
shown above, multiplying the coefficient
$\zeta(3)/(128\pi^4T^2)$, but these terms have been neglected. 
In fact, the renormalization structure
of the theory suggests the parametric convention 
$m\sim h_t T$, $u_s \sim h_t^2 T$, according 
to which all the terms involving $\zeta(3)/(128\pi^4)$
would be of higher order. Nevertheless, 
we keep the terms shown since numerically the mixing parameters
might be larger than some of the masses.

The $L_b(\bmu), L_f(\bmu)$-terms
on the first rows of eqs.~\nr{mm13}--\nr{mm123} 
cancel the running of $m_1^2(\bmu)$, 
$m_2^2(\bmu)$, $m_{12}^2(\bmu)$ so that the 3d masses 
are RG-invariant at 1-loop order. More precisely,
the effect 
of the $L_b(\bmu), L_f(\bmu)$-terms is to run 
the $\msbar$ mass parameters to a 
certain scale $\bmu_T$, which need not be the same
for all the parameters. 
For instance, if there are only bosonic contributions, 
then it can be seen from eq.~\nr{Lb} that $\bmu_T\approx 7T$.
In Sec.~\ref{vacren}
the running parameters $m_1^2(\bmu)$, 
$m_2^2(\bmu)$, $m_{12}^2(\bmu)$ are expressed in terms of
physical parameters and $\bmu$ so that the 
$\bmu$-dependence cancels in the 3d parameters.

Finally, the parameter $\xi$ in eq.~\nr{LHiggs} is 
\be
\xi=\frac{\zeta(3)}{128\pi^4T^2}(u_sw_s-d_se_s). \la{drxi}
\ee
If the $g^2$-corrections from the Higgs fields 
were included, there
would also be a term proportional to $g^2m_{12}^2$
inside the parentheses in~\nr{drxi}. 

3. The quadratic terms needed in the squark sector are
\ba
L_{\rm squark} & = & 
(D_iQ_\alpha)^\dagger(D_iQ_\alpha)+
(\partial_iU_\alpha)^*(\partial_iU_\alpha)+
(\partial_iD_\alpha)^*(\partial_iD_\alpha) \nn \\
& + & 
m_{Q3}^2 Q_\alpha^\dagger Q_\alpha+
m_{U3}^2 U_\alpha^* U_\alpha+
m_{D3}^2 D_\alpha^* D_\alpha. \la{Lsquark}
\ea
The graphs contributing to the 2-point correlator
$\langle Q^\dagger Q\rangle$ are
(AQ), (CQ), (SH), (A), (C), (S), (H) in Fig.~\ref{fig:dimred}.a; 
for $\langle U^*U\rangle$ and $\langle D^*D\rangle$ 
the interactions with 
SU(2) gauge fields are missing.
The fields in~\nr{Lsquark} are related to the original fields by 
\ba
(Q^\dagger Q)^{({\rm new})} & = & \frac{1}{T}
(Q^\dagger Q)(\bmu)\biggl[1-\pf 4 g_S^2 L_b(\bmu)+
\frac{\zeta(3)}{384\pi^4}\frac{u_s^2+d_s^2+w_s^2+e_s^2}{T^2}
\biggr], \hspace*{0.8cm}  \\
(U^*U)^{({\rm new})} & = & \frac{1}{T}
(U^*U)(\bmu)\biggl[1-\pf 4 g_S^2 L_b(\bmu)+
\frac{\zeta(3)}{192\pi^4}\frac{u_s^2+w_s^2}{T^2}
\biggr], \\
(D^*D)^{({\rm new})} & = & \frac{1}{T}
(D^*D)(\bmu)\biggl[1-\pf 4 g_S^2 L_b(\bmu)+
\frac{\zeta(3)}{192\pi^4}\frac{d_s^2+e_s^2}{T^2}
\biggr], 
\ea
where the $g^2$-terms have been neglected.
The terms proportional to $g_S^2$ represent
the contributions within the present theory, and 
are due to gluon loops.

The mass parameters in eq.~\nr{Lsquark} are
\ba
m_{Q3}^2 & = & m_Q^2(\bmu)-
\pf L_b(\bmu)\Bigl[
h_t^2(m_2^2+m_U^2)+h_b^2(m_1^2+m_D^2)\nn \\
& + & u_s^2+d_s^2+e_s^2+w_s^2
-\fr83 g_S^2m_Q^2
\Bigr] 
+\biggl(\fr14 g^2+\fr49 g_S^2+\fr16 h_t^2+
\fr16 h_b^2\biggr)T^2 \nn \\
& + & \frac{\zeta(3)}{128 \pi^4 T^2}
\Bigl[
u_s^2 (m_2^2+m_U^2)+
d_s^2 (m_1^2+m_D^2)+
w_s^2 (m_1^2+m_U^2)+
e_s^2 (m_2^2+m_D^2)\nn \\
& + & 
2 (u_s w_s-d_s e_s)m_{12}^2-
\fr13 (u_s^2+d_s^2+e_s^2+w_s^2)m_Q^2
\Bigr], \la{mmQ3} \\
m_{U3}^2 & = & m_U^2(\bmu)-
\pf L_b(\bmu)\Bigl[
2 h_t^2(m_2^2+m_Q^2)+2 (u_s^2+w_s^2)
-\fr83 g_S^2m_U^2
\Bigr]  \nn \\
& + & \biggl(\fr49 g_S^2+\fr13 h_t^2\biggr)T^2 
+\frac{\zeta(3)}{128 \pi^4 T^2}
\Bigl[
2 u_s^2 (m_2^2+m_Q^2)+
2 w_s^2 (m_1^2+m_Q^2) \nn \\
& + & 
4 u_s w_s m_{12}^2-
\fr23 (u_s^2+w_s^2)m_U^2
\Bigr], \la{mmU3} \\
m_{D3}^2 & = & m_D^2(\bmu)-
\pf L_b(\bmu)\Bigl[
2 h_b^2(m_1^2+m_Q^2)+2(d_s^2+e_s^2)
-\fr83 g_S^2m_D^2
\Bigr] \nn  \\
& + & \biggl(\fr49 g_S^2+
\fr13 h_b^2\biggr)T^2 
+\frac{\zeta(3)}{128 \pi^4 T^2}
\Bigl[
2 d_s^2 (m_1^2+m_Q^2)+
2 e_s^2 (m_2^2+m_Q^2) \nn \\ 
& - & 
4 d_s e_s m_{12}^2-
\fr23 (d_s^2+e_s^2)m_D^2
\Bigr].\la{mmD3}
\ea
The $L_b(\bmu)$-terms cancel 
the running of the soft masses 
$m_Q^2(\bmu), m_U^2(\bmu), m_D^2(\bmu)$ so that
the 3d masses, like the Higgs sector masses, 
are RG-invariant at 1-loop level.
It should be noted that the
running may be noticeable; for instance,
the parameter $m_U^2$ which may be small 
has a running proportional to $h_t^2m_Q^2$, so that
the relative effect may be significant. As in the 
case of Higgs mass parameters, one should hence 
renormalize the squark mass sector 
at the 1-loop level to remove 
the $\bmu$-dependence (1-loop
corrections to the stop mass have been calculated in~\cite{don}).
However, since the squark masses 
at zero temperature are not known at the moment, we will 
not perform any renormalization in the present investigation.
Instead, the parameters $m_Q^2(\bmu_T)$, $m_U^2(\bmu_T)$, 
$m_D^2(\bmu_T)$
produced by the $L_b(\bmu)$-terms in eqs.~\nr{mmQ3}-\nr{mmD3}
are replaced by the tree-level values.

4. The interactions of $A_0^a$ and $C_0^A$ with the Higgs fields
are produced by the graphs (qqqq), (SS), (SSS1), (SSS2), (SSSS)
in Fig.~\ref{fig:dimred}.c.
There are terms of the form 
\be
L^{\rm tree}=
h_1 A_0^aA_0^a H^{1\dagger}H^1+
h_2 A_0^aA_0^a H^{2\dagger}H^2 
\ee
existing already at the tree-level, 
and terms generated radiatively, 
\ba
L^{\rm rad} & = & 
h_3 A_0^aA_0^a (\hahb+\hbha)  \\
& + &  c_1 C_0^AC_0^A \haha+
c_2 C_0^AC_0^A \hbhb+
c_3 C_0^AC_0^A (\hahb+\hbha). \nn
\ea
The coefficients related to $A_0^a$ are
\ba
h_1 & = & \frac{g_3^2}{4}-\frac{g^2\zeta(3)}{128\pi^4}
\frac{d_s^2+w_s^2}{T}, \la{h1} \\
h_2 & = & \frac{g_3^2}{4}-\frac{g^2\zeta(3)}{128\pi^4}
\frac{u_s^2+e_s^2}{T}, \la{h2} \\
h_3 & = & -\fr34\frac{g^2 \zeta(3)}{128\pi^4}\frac{u_s w_s-d_s e_s}{T}. 
\la{h3}
\ea
Note that in the Standard Model 
extra terms of the type $g^2h_t^2T/(16\pi^2)$ 
are generated in $h_1, h_2$ through quark 
loops~\cite{klrs1}, but in MSSM these terms
are cancelled by the squark loops.
If the $g^2$-corrections from Higgs fields were included, there
would be a term proportional to $g^2m_{12}^2$
in eq.~\nr{h3}.

As to the coefficients $c_1, c_2$, the quark 
loops (qqqq) give the contributions 
\be
\delta c_1=-\frac{T}{8\pi^2} g_S^2h_b^2,\quad
\delta c_2=-\frac{T}{8\pi^2} g_S^2h_t^2
\ee
as in the Standard Model, but these are
cancelled by the squark loops (SS), (SSS1)
in Fig.~\ref{fig:dimred}.c,
as for $h_1, h_2$. Hence
there only remain the small terms
\ba
c_1 & = & -\frac{g_S^2\zeta(3)}{64\pi^4}\frac{d_s^2+w_s^2}{T},\\
c_2 & = & -\frac{g_S^2\zeta(3)}{64\pi^4}\frac{u_s^2+e_s^2}{T},\\
c_3 & = & -\frac{g_S^2\zeta(3)}{64\pi^4}
\frac{u_s w_s-d_s e_s}{T}.
\ea

5. For the quartic self-interactions of the Higgs fields, the most 
general gauge-invariant two Higgs doublet potential~\cite{hh}
is generated at the dimensional reduction step.
Since we assumed all the parameters to be real in 
the original Lagrangian, the potential is somewhat 
simplified, being of 
the form\footnote{We recall that the identity
$(H^{1\dagger}H^2)(H^{2\dagger}H^1)+(\hahb)(\hbha) = (\haha)(\hbhb)$
reduces the number of independent combinations.}
\ba
V\!\! & = & \!\! 
\lambda_1 (\haha)^2+\lambda_2 (\hbhb)^2+
\lambda_3 \haha\hbhb+\lambda_4 \hahb \hbha \\
& + & \!\! \lambda_5(\hahb\hahb+\hc)+
\lambda_6\haha(\hahb+\hc)+
\lambda_7\hbhb(\hahb+\hc). \nn 
\ea
To give the expressions for $\lambda_1,\ldots,\lambda_7$, 
we use the functions
\ba
f_2(m_a,m_b) & \equiv & \Tint{p_b}'\biggl[
\frac{1}{(p^2+m_a^2)(p^2+m_b^2)}-\frac{1}{(p^2)^2}
\biggr] \la{f2ab} \\
& =  & -\frac{\zeta(3)}{128\pi^4}\frac{m_a^2+m_b^2}{T^2}+
{\cal O}\biggl(\frac{m^4}{16\pi^2(2\pi T)^4}\biggr) , \nn \\
f_3(m_a,m_b,m_c) & \equiv & \Tint{p_b}'
\frac{1}{(p^2+m_a^2)(p^2+m_b^2)(p^2+m_c^2)}  \\
& = & 
\frac{\zeta(3)}{128\pi^4T^2}
-\frac{\zeta(5)}{1024\pi^6}\frac{m_a^2+m_b^2+m_c^2}{T^4}+
{\cal O}\biggl(\frac{m^4}{16\pi^2(2\pi T)^6}\biggr) , \nn \\
f_4(m_a,m_b,m_c,m_d) & \equiv & \Tint{p_b}'
\frac{1}{(p^2+m_a^2)(p^2+m_b^2)(p^2+m_c^2)(p^2+m_d^2)} \la{f4} \\
& = &  \frac{\zeta(5)}{1024\pi^6T^4}+
{\cal O}\biggl(\frac{m^2}{16\pi^2 (2\pi T)^6}\biggr), \nn
\ea
where the sum-integral is
over the non-zero bosonic Matsubara frequencies in the $\msbar$-scheme. 
In the numerical computations
we keep only the constant 
part in the function $f_3$, to be consistent with
the fact that terms of the same parametric form 
come from the redefinition of fields and 
the higher mass contributions were there neglected. The part subtracted 
in the definition of $f_2(m_a,m_b)$ is 
\be
\Tint{p_b}'\frac{1}{(p^2)^2}=\pf
\biggl[\frac{1}{\epsilon}+L_b(\bmu)\biggr].
\ee

The graphs needed are (qqqq), (SS), (SSS), (SSSS)
in Fig.~\ref{fig:dimred}.b. In addition, the redefinitions
of fields according to eqs.~\nr{haha}--\nr{hbhb}
give contributions. After the redefinition, 
the parameters are ($\hat{g}_3^2\equiv g_3^2/T$):
\ba
\frac{\lambda_1}{T} \!\! & = & \!\!
\frac{g'^2}{8} + \frac{\hat{g}_3^2}{8}
+\pf \Bigl[L_b(\bmu)-L_f(\bmu)\Bigr]
\Bigl(
-\fr18 g^4
- 3 h_b^4 
+\fr34 g^2 h_b^2 \Bigr) \la{lam1} \\
& - & \!\! \fr14 g^2 (d_s^2+w_s^2)\frac{\zeta(3)}{128\pi^4T^2}
\nn \\
& + & \!\! \biggl\{-\frac{3}{16}g^4 f_2(m_Q,m_Q) 
- \fr32 h_b^4 \Bigl[f_2(m_Q,m_Q)+f_2(m_D,m_D)\Bigr]
+\fr34 g^2 h_b^2 
f_2(m_Q,m_Q) \nn \\
& + & \!\!
3 h_b^2 d_s^2 \Bigl[f_3(m_D,m_D,m_Q)+f_3(m_Q,m_Q,m_D)\Bigr] 
 +  \fr34 g^2 w_s^2 f_3(m_Q,m_Q,m_U) \nn \\
& - &\!\!\!\! \fr34 g^2 d_s^2 f_3(m_Q,m_Q,m_D)\! 
-\fr32 d_s^4 f_4(m_Q,m_Q,m_D,m_D)
-\fr32 w_s^4 f_4(m_Q,m_Q,m_U,m_U)\biggr\}, \nn \\
\frac{\lambda_2}{T} \!\! & = &  \!\!
\frac{g'^2}{8}+\frac{\hat{g}_3^2}{8}
+\pf \Bigl[L_b(\bmu)-L_f(\bmu)\Bigr]
\Bigl(
-\fr18 g^4
- 3 h_t^4
+\fr34 g^2 h_t^2 \Bigr) \la{lam2} \\
& - & \!\! \fr14 g^2 (u_s^2+e_s^2) \cTT 
\nn \\
& + & \!\! \biggl\{
-\frac{3}{16}g^4 f_2(m_Q,m_Q)
- \fr32 h_t^4 \Bigl[f_2(m_Q,m_Q)+f_2(m_U,m_U)\Bigr]
+\fr34 g^2 h_t^2 
f_2(m_Q,m_Q)
\nn \\
& + & \!\! 3 h_t^2 u_s^2 \Bigl[f_3(m_U,m_U,m_Q)+f_3(m_Q,m_Q,m_U)\Bigr] 
-\fr34 g^2 u_s^2 f_3(m_Q,m_Q,m_U) \nn \\
& + &\!\!\!\! \fr34 g^2 e_s^2 f_3(m_Q,m_Q,m_D)\!
- \fr32 u_s^4 f_4(m_Q,m_Q,m_U,m_U)
-\fr32 e_s^4 f_4(m_Q,m_Q,m_D,m_D) \biggr\}, \nn \\
\frac{\lambda_3}{T} & = & 
-\frac{g'^2}{4}+\frac{\hat{g}_3^2}{4}
+\pf \Bigl[L_b(\bmu)-L_f(\bmu)\Bigr]\Bigl[
-\fr14 g^4 
- 6 h_t^2 h_b^2
+\fr34 g^2 (h_t^2+h_b^2)\Bigr] \la{lam3} \\
& - & \fr14 g^2 (u_s^2+d_s^2+e_s^2+w_s^2) \cTT
+ \biggl\{-\frac{3}{8}g^4 
f_2(m_Q,m_Q) \nn \\
& - & 3 h_t^2 h_b^2\Bigl[
f_2(m_Q,m_Q)+f_2(m_U,m_D)\Bigr]
+\fr34 g^2 (h_t^2+h_b^2)f_2(m_Q,m_Q) \nn \\
& + & 3 h_t^2 w_s^2 f_3(m_U,m_U,m_Q) 
+3 h_t^2 d_s^2 f_3(m_Q,m_Q,m_D) 
+3 h_b^2 e_s^2 f_3(m_D,m_D,m_Q) \nn \\
& +  &3 h_b^2 u_s^2 f_3(m_Q,m_Q,m_U) 
+6 h_t h_b (u_s d_s+e_s w_s) f_3(m_Q,m_U,m_D) \nn \\
& + & \fr34 g^2 (w_s^2-u_s^2) f_3(m_Q,m_Q,m_U) 
+\fr34 g^2(e_s^2-d_s^2) f_3(m_Q,m_Q,m_D) \nn \\
& - & 3 (u_s d_s+e_s w_s)^2 f_4(m_Q,m_Q,m_U,m_D) 
-3 u_s^2 w_s^2 f_4(m_Q,m_Q,m_U,m_U) \nn \\
& - & 3 d_s^2 e_s^2 f_4(m_Q,m_Q,m_D,m_D)\biggr\}, \nn \\
\frac{\lambda_4}{T} & = &       
-\frac{\hat{g}_3^2}{2}
+\pf \Bigl[L_b(\bmu)-L_f(\bmu)\Bigr]\Bigl[
\fr12 g^4 
+ 6 h_t^2 h_b^2 
-\fr32 g^2 (h_t^2+h_b^2)
\Bigr] \la{lam4} \\
& + & \fr12 g^2 (u_s^2+d_s^2+e_s^2+w_s^2)\cTT
+ \biggl\{\frac{3}{4}g^4 f_2(m_Q,m_Q)
\nn \\
& + & 3 h_t^2 h_b^2 \Bigl[
f_2(m_Q,m_Q)+f_2(m_U,m_D)\Bigr]
-\fr32 g^2 (h_t^2+h_b^2)
f_2(m_Q,m_Q) \nn \\
& + & 3 h_t^2 w_s^2 f_3(m_Q,m_Q,m_U)
-3 h_t^2 d_s^2 f_3(m_Q,m_Q,m_D)
+3 h_b^2 e_s^2 f_3(m_Q,m_Q,m_D) \nn \\
& - & 3 h_b^2 u_s^2 f_3(m_Q,m_Q,m_U)
-6 h_t h_b (u_s d_s+e_s w_s) f_3(m_Q,m_U,m_D) \nn \\
& +  & \fr32 g^2 (u_s^2-w_s^2) f_3(m_Q,m_Q,m_U)
+\fr32 g^2 (d_s^2-e_s^2) f_3(m_Q,m_Q,m_D) \nn \\
& +  &3 (u_s d_s+e_s w_s)^2 f_4(m_Q,m_Q,m_U,m_D)
-3 u_s^2 w_s^2 f_4(m_Q,m_Q,m_U,m_U) \nn \\
& - & 3 d_s^2 e_s^2 f_4(m_Q,m_Q,m_D,m_D)\biggr\}, \nn \\ 
\frac{\lambda_5}{T} & = & \biggl\{
-\fr32 u_s^2 w_s^2 f_4(m_Q,m_Q,m_U,m_U)
-\fr32 d_s^2 e_s^2 f_4(m_Q,m_Q,m_D,m_D)\biggl\}, \\
\frac{\lambda_6}{T} & = & \biggl\{
-3 h_b^2 d_s e_s \Bigl[f_3(m_D,m_D,m_Q)+f_3(m_Q,m_Q,m_D)\Bigr] \la{lam6} \\
& + & \fr34 g^2 u_s w_s f_3(m_Q,m_Q,m_U) 
+ \fr34 g^2 d_s e_s f_3(m_Q,m_Q,m_D) \nn \\
& - & 3 u_s w_s^3 f_4(m_Q,m_Q,m_U,m_U)
+3 e_s d_s^3 f_4(m_Q,m_Q,m_D,m_D)\biggl\}, \nn \\
\frac{\lambda_7}{T} & = & \biggl\{
3 h_t^2 u_s w_s \Bigl[f_3(m_U,m_U,m_Q)+f_3(m_Q,m_Q,m_U)\Bigl]\la{lam7} \\
& - & \fr34 g^2 u_s w_s f_3(m_Q,m_Q,m_U) 
-\fr34 g^2 d_s e_s f_3(m_Q,m_Q,m_D) \nn \\
& - & 3 w_s u_s^3 f_4(m_Q,m_Q,m_U,m_U) 
+3 d_s e_s^3 f_4(m_Q,m_Q,m_D,m_D)\biggl\}. \nn
\ea
The terms in the curly brackets will be useful in
Sec.~\ref{squarks} as well, which is why they have been
separated.

6. Cubic interactions of Higgs fields and squarks are
of the same form as in the original theory:
\be
L_{\rm cubic}=
u_{s3}\thbd Q_\alpha U_\alpha+
d_{s3}\thad Q_\alpha D_\alpha+
e_{s3}\hbd Q_\alpha D_\alpha+
w_{s3}\had Q_\alpha U_\alpha+\hc.
\ee 
Since the terms $u_s, d_s, w_s, e_s$ are unknown, 
it is not so important at the moment to calculate
the 1-loop corrections to the tree-level formulas. Just as an illustration
of the structure that appears,  
let us give the 1-loop terms proportional to 
$h_t, h_b$ within the present theory:
\ba
\frac{u_{s3}}{\sqrt{T}} & = & 
u_s(\bmu)-\pf \Bigl[(6 h_t^2 u_s+h_th_b d_s)L_b(\bmu)
+ \fr32 h_t^2 u_s  L_f(\bmu)\Bigr], \\
\frac{d_{s3}}{\sqrt{T}} & = & 
d_s(\bmu)-\pf \Bigl[(6 h_b^2 d_s+h_th_b u_s)L_b(\bmu)
+  \fr32 h_b^2 d_s L_f(\bmu)\Bigr], \\
\frac{e_{s3}}{\sqrt{T}} & = & 
e_s(\bmu)-\pf \Bigl[(3 h_b^2 e_s-h_t^2 e_s+h_th_bw_s)L_b(\bmu)
+  \fr32 h_t^2 e_s L_f(\bmu)\Bigr], \\
\frac{w_{s3}}{\sqrt{T}} & = & 
w_s(\bmu)-\pf \Bigl[(3 h_t^2 w_s-h_b^2 w_s+h_th_be_s)L_b(\bmu)
+ \fr32 h_b^2 w_s  L_f(\bmu)\Bigr]. 
\ea

7. Quartic interactions of Higgs fields and squarks
are at tree-level of the form
\ba
L_{\rm quartic} & = & 
\fr14 g_3^2 H^{m*}_iH^m_jQ^*_{k\alpha}Q_{l\alpha}
(2\delta_{il}\delta_{jk}-\delta_{ij}\delta_{kl}) \nn \\
& + & 
h_{t3}^2(\hbd H^2 U^*_\alpha U_\alpha+
\thbd Q_\alpha Q^\dagger_\alpha\tilde{H}^2)+
h_{b3}^2(\had H^1 D^*_\alpha D_\alpha+
\thad Q_\alpha Q^\dagger_\alpha\tilde{H}^1) \nn \\
& + &  h_{t3}h_{b3}
(H^{2\dagger}H^1 U^*_\alpha D_\alpha +\hc). \la{Lquartic}
\ea
In principle it would be important to calculate the 
1-loop corrections especially to $h_{t3}$ since it affects
the transition quite significantly. However, as stated above, 
this is
not accessible within the present framework, 
since vacuum renormalization of the top quark mass 
cannot be used to simultaneously fix the $h_t$'s appearing in 
different places in the Lagrangian beyond tree-level.  
Moreover, the top mass is not known very accurately, 
and the effect of $h_{t3}$ comes 
together with $m_{U3}$  which is not known at all.
Hence we take the couplings here only at tree-level.
Then the couplings are as written in eq.~\nr{Lquartic} with
\be
h_{t3}^2=h_t^2 T,\quad
h_{b3}^2=h_b^2 T.
\ee

8. Finally, there are many terms not interacting 
directly with the SU(2) gauge fields and Higgs fields.
We will not show them explicitly, since they enter the
further integrations only at 2-loop level. Nevertheless, 
in some cases the higher-order corrections are 
important; a particularly relevant example~\cite{e} is discussed
in Sec.~\ref{Utheory}. We just fix one more notation here:
the strong coupling constant in 3d is denoted by $g_{S3}^2$ 
and is $g_{S3}^2=g_S^2(\bmu_T)T$. 

\section{Integrating out squarks, $A^a_0$ and $C^A_0$}
\la{squarks}

The bosonic theory discussed in Sec.~\ref{dimred}
is still rather complicated, although simpler than the original 
theory. However, generically many of the fields appearing
are massive at the phase transition point. 
Such fields can be integrated out in 3d. It appears that in some
part of the parameter space, all the squarks together with 
the adjoint scalar fields $A_0^a, C_0^A$ can be integrated out.

More specifically, the requirements for the integration
to be valid are the following. First, the phase transition 
should be weak enough so that the neglected higher-order operators
are not important. Second, the perturbative expansion for the 
parameters of the effective theory should converge. The first 
requirement should be reasonably well satisfied when $x \gsim 0.03$ 
which is the region we are studying. Let us investigate
the second requirement in some more detail.

Integrating out $A_0^a, C_0^A$ gives roughly the expansion parameters
\be
\frac{g_3^2}{4\pi m_{A_0}}, \quad
\frac{g_{S3}^2}{4\pi m_{C_0}}. \la{a0exp}
\ee
{}From eqs.~\nr{ma0}, \nr{mc0} one sees that these are small 
numbers, below 0.05 (to be more precise, the expansion
parameter of $C_0^A$-integration might be slightly larger 
due to colour factors, but on the other
hand $g_{S3}^2$ appears first only at 2-loop level). 
With the trilinear couplings
(which have the dimension GeV$^{3/2}$ in 3d)
are associated expansion parameters of the type
\be
\frac{u_{s3}^2}{4\pi m_{Q3}^3},
\ee
which are very small for small mixing. The largest 
and most important expansion parameters are related to 
the strongly interacting squarks. There the expansion
proceeds in powers of (see Sec.~\ref{Utheory}) 
\be
\frac{g_{S3}^2}{\pi m_{U3}},
\frac{h_{t3}^2}{\pi m_{U3}} \la{expU3}
\ee
and correspondingly for the other squarks.
Roughly, the factor 4 in the denominator of~\nr{a0exp} is
compensated in~\nr{expU3} by colour factors.
The terms in eq.~\nr{expU3} are of order 0.3 if
\be 
m_{U3}^2 \sim m_U^2+(4g_S^2/9+h_t^2/3) T^2 \sim T^2, 
\ee
in which case the neglected 2-loop terms
are expected to give a correction
of about 20\% (1-loop corrections may sometimes be 
almost as large as tree-level terms). 
In the present Section we shall 
assume that $m_U\gsim 50$ GeV so that the expansion
in~\nr{expU3} 
should still be useful (for the other squarks, 
we assume $m_Q \sim m_D \sim 300$ GeV).
The case of smaller $m_U$
is discussed in Sec.~\ref{Utheory}.

In the case that all the squarks and the $A_0^a$- and 
$C_0^A$-fields can be integrated out, the new theory 
will be 
\ba
L\!\!  & = &\!\! \fr14 F^a_{ij}F^a_{ij}
+(D_iH^1)^\dagger(D_iH^1)+
(D_iH^2)^\dagger(D_iH^2)+
\xi \Bigl[
(D_iH^1)^\dagger(D_i\tilde{H}^2)+\hc
\Bigr] \la{L3d2hd} \\
& + & \!\! m_1^2 H^{1\dagger}H^1+
m_2^2 H^{2\dagger}H^2+m_{12}^2 (H^{1\dagger}\tilde{H}^2+
\tilde{H}^{2\dagger}H^1) \nn \\
& + & \!\!
\lambda_1 (\haha)^2+\lambda_2 (\hbhb)^2+
\lambda_3 \haha\hbhb+\lambda_4 \hahb \hbha \nn \\
& + &\!\! \lambda_5(\hahb\hahb+\hc)+
\lambda_6\haha(\hahb+\hc)+
\lambda_7\hbhb(\hahb+\hc). \nn 
\ea
Although the notation for the parameters is the same as before,
the parameters have changed from the previous theory.

The graphs needed for calculating the parameters have
a simple relation to the graphs needed in the dimensional reduction
step. The quark contributions do not exist any more. The
squark graphs remain precisely the same. In addition, there
are the extra graphs with $A_0^a$, $C_0^A$
in the internal lines, of the same type as for squarks but
without cubic interactions with the Higgs fields.

{}From the graphs (SS), ($A_0A_0$) in 
Fig.~\ref{fig:dimred}.a, one gets that the 
new fields are related to the previous ones by
\ba
\Bigl(A^a_iA^b_j\Bigr)^{({\rm new})} & = & 
\Bigl(A^a_iA^b_j\Bigr) \biggl[
1+\frac{g_3^2}{16\pi m_{Q3}}+\frac{g_3^2}{24\pi m_{A_0}}
\biggr], \\ 
\Bigl(H^{1\dagger}H^1\Bigr)^{({\rm new})} & = & 
\Bigl(H^{1\dagger}H^1\Bigr) Z_{H1}^2 \\
\Bigl(H^{2\dagger}H^2\Bigr)^{({\rm new})} & = &  
\Bigl(H^{2\dagger}H^2\Bigr) Z_{H2}^2, 
\ea
where
\ba
Z_{H1}^2 & = & 1+\frac{1}{4\pi}\biggl(
\frac{w_{s3}^2}{(m_{Q3}+m_{U3})^3}+
\frac{d_{s3}^2}{(m_{Q3}+m_{D3})^3}
\biggr), \la{Zh1} \\ 
Z_{H2}^2 & = &
1+\frac{1}{4\pi}\biggl(
\frac{u_{s3}^2}{(m_{Q3}+m_{U3})^3}+
\frac{e_{s3}^2}{(m_{Q3}+m_{D3})^3}
\biggr). \la{Zh2}
\ea
The parameter $\xi$ is changed to be
\be
\xi^{({\rm new})} =
\xi+\frac{1}{4\pi}\biggl(
\frac{u_{s3}w_{s3}}{(m_{Q3}+m_{U3})^3}-
\frac{d_{s3}e_{s3}}{(m_{Q3}+m_{D3})^3}
\biggr).
\ee
The new gauge coupling can be derived from the 
graphs ($A_0A_0$), ($A_0A_0A_01$), (SS), (SSS1), (SSS2), (SSSS)
in Fig.~\ref{fig:dimred}.c and is
\be
g_3^{2(\rm new)} = 
g_3^2 \biggl[
1-\frac{g_3^2}{16\pi m_{Q3}}-\frac{g_3^2}{24\pi m_{A_0}}
\biggr], 
\ee
so that $g_3^{2(\rm new)}\Bigl(A^a_iA^b_j\Bigr)^{({\rm new})}=
g_3^2\Bigl(A^a_iA^b_j\Bigr)$.
The new mass parameters are given by
\ba
m_1^{2{({\rm new})}}Z_{H1}^2 & = & 
m_1^2-\frac{3}{4\pi} h_1 m_{A_0}
-\frac{8}{4\pi} c_1 m_{C_0}  \\
& - & \frac{3}{4\pi} h_{b3}^2 
(m_{Q3}+m_{D3})
-\frac{3}{4\pi}
\biggl(
\frac{w_{s3}^2}{m_{Q3}+m_{U3}}+
\frac{d_{s3}^2}{m_{Q3}+m_{D3}}
\biggr), \nn \\
m_2^{2{({\rm new})}}Z_{H2}^2 & = & 
m_2^2-\frac{3}{4\pi} h_2 m_{A_0}
-\frac{8}{4\pi} c_2 m_{C_0} \\
& - & \frac{3}{4\pi} h_{t3}^2 
(m_{Q3}+m_{U3}) 
-\frac{3}{4\pi}
\biggl(
\frac{u_{s3}^2}{m_{Q3}+m_{U3}}+
\frac{e_{s3}^2}{m_{Q3}+m_{D3}}
\biggr), \nn \\
m_{12}^{2{({\rm new})}}Z_{H1}Z_{H2} & = & 
m_{12}^2 
-\frac{3}{4\pi} h_3 m_{A_0}
-\frac{8}{4\pi} c_3 m_{C_0}  \\
& - & \frac{3}{4\pi}
\biggl(
\frac{u_{s3}w_{s3}}{m_{Q3}+m_{U3}}-
\frac{d_{s3}e_{s3}}{m_{Q3}+m_{D3}}
\biggr). \nn
\ea

For the scalar coupling constants, one can to a large extent use 
the results in eqs.~\nr{lam1}--\nr{lam7}.
The squark graphs and the combinatorial factors are 
precisely the same, but the integration measure and 
the parameters appearing have changed. The fermion graphs 
are missing, but the $A_0, C_0$-graphs have to be included.
Hence the graphs are 
(SS), (SSS), (SSSS), ($A_0A_0$), ($C_0C_0$) in Fig.~\ref{fig:dimred}.b. 
We will display only the part
arising from $A_0^a, C_0^A$ explicitly; the rest
can be read from eqs.~\nr{lam1}--\nr{lam7} and is 
indicated by the curly brackets below.
The replacements to be made in~\nr{lam1}--\nr{lam7} are that 
the functions $f_2, f_3, f_4$ are replaced with those
defined in eqs.~\nr{f23d}--\nr{f43d} below; 
$g, h_t, u_s, \ldots \to$ 
$g_3, h_{t3}, u_{s3},\ldots$; and 
$m_Q, m_U, m_D \to$
$m_{Q3}, m_{U3}, m_{D3}$.

The integrals appearing, analogously to~\nr{f2ab}--\nr{f4}, are
\ba
f_2(m_a,m_b) & \equiv & \int dp
\frac{1}{(p^2+m_a^2)(p^2+m_b^2)}=\frac{1}{4\pi(m_a+m_b)}, \la{f23d}\\
f_3(m_a,m_b,m_c) & \equiv & \int dp
\frac{1}{(p^2+m_a^2)(p^2+m_b^2)(p^2+m_c^2)} \nn \\ 
& = & \frac{1}{4\pi(m_a+m_b)(m_a+m_c)(m_b+m_c)}, \\
f_4(m_a,m_b,m_c,m_d) & \equiv & \int dp
\frac{1}{(p^2+m_a^2)(p^2+m_b^2)(p^2+m_c^2)(p^2+m_d^2)} \nn
\ea
\be
\hspace*{1cm} =\frac{1}{4\pi}\frac{m_a+m_b+m_c+m_d}
{(m_a+m_b)(m_a+m_c)(m_a+m_d)(m_b+m_c)(m_b+m_d)(m_c+m_d)}. \la{f43d} 
\ee
The integration measure here is 
\be
\int dp \equiv \int \frac{d^dp}{(2\pi)^d}, \quad d=3-2\epsilon.
\ee
There is no divergence in $f_2(m_a,m_b)$ in 3d, so that 
it was not necessary to subtract anything in the definition 
in contrary to the 4d case. 

With the notation introduced, 
the new parameters are
\ba
\lambda_1^{({\rm new})} Z_{H1}^4 \!\! & = & \!\! \lambda_1 
-3 h_1^2 f_2(m_{A_0},m_{A_0})
-8 c_1^2 f_2(m_{C_0},m_{C_0}) 
+\Bigl\{ f_2-f_4 \Bigr\},\hspace*{1.0cm} \\
\lambda_2^{({\rm new})} Z_{H2}^4 \!\! & = & \!\! \lambda_2 
-3 h_2^2 f_2(m_{A_0},m_{A_0})
-8 c_2^2 f_2(m_{C_0},m_{C_0}) 
+\Bigl\{ f_2-f_4 \Bigr\}, \la{alam2} \\
\lambda_3^{({\rm new})} Z_{H1}^2 Z_{H2}^2 \!\! & = & \!\! \lambda_3 
-6 h_1 h_2 f_2(m_{A_0},m_{A_0})
-16 c_1 c_2 f_2(m_{C_0},m_{C_0}) 
+\Bigl\{ f_2-f_4 \Bigr\}, \\
\lambda_4^{({\rm new})} Z_{H1}^2 Z_{H2}^2 \!\! & = & \!\! \lambda_4 
-6 h_3^2 f_2(m_{A_0},m_{A_0})
-16 c_3^2 f_2(m_{C_0},m_{C_0}) 
+\Bigl\{ f_2-f_4 \Bigr\}, \\
\lambda_5^{({\rm new})} Z_{H1}^2 Z_{H2}^2 \!\! & = & \!\! \lambda_5 
-3 h_3^2 f_2(m_{A_0},m_{A_0})
-8 c_3^2 f_2(m_{C_0},m_{C_0}) 
+\Bigl\{ f_2-f_4 \Bigr\}, \\
\lambda_6^{({\rm new})} Z_{H1}^3 Z_{H2} \!\! & = & \!\! \lambda_6 
-6 h_1 h_3 f_2(m_{A_0},m_{A_0})
-16 c_1 c_3 f_2(m_{C_0},m_{C_0}) 
+\Bigl\{ f_2-f_4 \Bigr\}, \\
\lambda_7^{({\rm new})} Z_{H1} Z_{H2}^3 \!\! & = & \!\! \lambda_7 
-6 h_2 h_3 f_2(m_{A_0},m_{A_0})
-16 c_2 c_3 f_2(m_{C_0},m_{C_0}) 
+\Bigl\{ f_2-f_4 \Bigr\}. \hspace*{1.0cm} \la{alam7}
\ea
Here
we have displayed the terms arising from field redefinitions
on the LHS of the formulas. The factors
$Z_{H1}$, $Z_{H2}$ are given in eqs.~\nr{Zh1}, \nr{Zh2}.

\section{Diagonalization of the two Higgs doublet model}
\la{diag}

The theory in eq.~\nr{L3d2hd} can still be simplified.
The phase transition should take place close to the 
point where the mass matrix has a zero eigenvalue. Then generically
the other mass is heavy. Recall
that at tree-level the sum of the 
eigenvalues of the mass matrix is $m_1^2+m_2^2=m_A^2$, 
and at finite temperature one gets positive thermal 
corrections to the masses. Hence one may integrate out the heavier
Higgs doublet as well.
In order to do so, we first diagonalize the two Higgs doublet model.

We make the diagonalization in two steps. In the first part
we rotate and rescale the fields so that the term 
\be
\xi \Bigl[
(D_iH^1)^\dagger(D_i\tilde{H}^2)+\hc
\Bigr] \la{diag1}
\ee
disappears from the Lagrangian in eq.~\nr{L3d2hd}. In the second part
we rotate the resulting fields so that the non-diagonal mass term
\be
m_{12}^2 (H^{1\dagger}\tilde{H}^2+
\tilde{H}^{2\dagger}H^1)  \la{diag2}
\ee
disappears. Then the resulting theory will be 
\ba
L & = & \fr14 F^a_{ij}F^a_{ij} 
+ (D_i\phi)^\dagger(D_i\phi)+
(D_i\theta)^\dagger(D_i\theta)
+ m_\phi^2 \phi^{\dagger}\phi+
m_\theta^2 \theta^{\dagger}\theta  \nn \\
& + & 
\lambda_1 (\phi^\dagger\phi)^2+\lambda_2 (\theta^\dagger\theta)^2+
\lambda_3 \phi^\dagger\phi\theta^\dagger\theta
+\lambda_4 \phi^\dagger\theta \theta^\dagger\phi \nn \\
& + & \lambda_5(\phi^\dagger\theta\phi^\dagger\theta+\hc)+
\lambda_6\phi^\dagger\phi(\phi^\dagger\theta+\hc)+
\lambda_7\theta^\dagger\theta(\phi^\dagger\theta+\hc). \la{phithe}
\ea
It should be noted that for small values of the 
squark mixing parameters, $\xi$ in~\nr{diag1} is very small so that 
the first part of the diagonalization is 
numerically inessential.

The first part of diagonalization proceeds by writing
\ba
H^1 & = & \frac{1}{\sqrt{2}}
\Bigl[
(1-\xi)^{-1/2}H^{1({\rm new})}+
(1+\xi)^{-1/2}\tilde{H}^{2({\rm new})}
\Bigr], \\
\tilde{H}^2 & = & \frac{1}{\sqrt{2}}
\Bigl[
-(1-\xi)^{-1/2}H^{1({\rm new})}+
(1+\xi)^{-1/2}\tilde{H}^{2({\rm new})}
\Bigr]. 
\ea
Expressed in terms of the new fields, the term in 
eq.~\nr{diag1} vanishes. The other parameters become 
\ba
m_1^{2({\rm new})} & = & 
(1-\xi)^{-1}(m_1^2+m_2^2-2 m_{12}^2)/2, \la{d1m1} \\
m_2^{2({\rm new})} & = & 
(1+\xi)^{-1}(m_1^2+m_2^2+2 m_{12}^2)/2, \\
m_{12}^{2({\rm new})} & = & 
(1-\xi^2)^{-1/2}(m_1^2-m_2^2)/2, \la{d1m12} \\
\lambda_1^{({\rm new})} & = & 
(1-\xi)^{-2}
(\lambda_1+\lambda_2+\lambda_3+\lambda_4+
2\lambda_5-2\lambda_6-2\lambda_7)/4, \la{dl1}\\
\lambda_2^{({\rm new})} & = & 
(1+\xi)^{-2}
(\lambda_1+\lambda_2+\lambda_3+\lambda_4+
2\lambda_5+2\lambda_6+2\lambda_7)/4, \\
\lambda_3^{({\rm new})} & = & 
(1-\xi^2)^{-1}
(\lambda_1+\lambda_2+\lambda_3-\lambda_4-
2\lambda_5)/2, \\
\lambda_4^{({\rm new})} & = & 
(1-\xi^2)^{-1}
(\lambda_1+\lambda_2-\lambda_3+\lambda_4-
2\lambda_5)/2, \\
\lambda_5^{({\rm new})} & = & 
(1-\xi^2)^{-1}
(\lambda_1+\lambda_2-\lambda_3-\lambda_4+
2\lambda_5)/4, \\
\lambda_6^{({\rm new})} & = & 
(1-\xi)^{-1}(1-\xi^2)^{-1/2}
(\lambda_1-\lambda_2-\lambda_6+\lambda_7)/2, \\
\lambda_7^{({\rm new})} & = & 
(1+\xi)^{-1}(1-\xi^2)^{-1/2}
(\lambda_1-\lambda_2+\lambda_6-\lambda_7)/2. \la{dl7}
\ea

In the second part of the diagonalization, we write
\ba
H^1 & = & \cos\!\alpha\,\phi+\sin\!\alpha\,\theta, \la{rot1} \\
\tilde{H}^2 & = & -\sin\!\alpha\,\phi+\cos\!\alpha\,\theta. \la{rot2}
\ea
The angle $\alpha$ is chosen so that 
\be
\tan\! 2\alpha=\frac{2 m_{12}^2}{m_2^2-m_1^2},\quad
\sin\! 2\alpha=\frac{2 m_{12}^2}
{\sqrt{(m_1^2-m_2^2)^2+4 m_{12}^4}}. \la{angle}
\ee
It should be reiterated that at 1-loop level 
in the $\msbar$ scheme the 3d mass parameters are finite, 
so that we need not worry about renormalization
at this point.
As a result of the rotation
in eqs.~\nr{rot1}, \nr{rot2}, 
the action is of the form in~\nr{phithe}.
The new mass parameters, obtained from 
those in~\nr{d1m1}--\nr{d1m12}, are
\ba
m_\phi^2 & = & \fr12
\Bigl[m_1^2+m_2^2-\sqrt{(m_1^2-m_2^2)^2+4 m_{12}^4}
\Bigr], \la{d2m1} \\
m_\theta^2 & = & \fr12
\Bigl[m_1^2+m_2^2+\sqrt{(m_1^2-m_2^2)^2+4 m_{12}^4}
\Bigr].
\ea
Abbreviating
$\cca=c^2, \ssa=s^2, \caa=c2, \saa=s2$, the matrix $M$
giving the couplings as $\lambda^{({\rm new})}=M\lambda$ from
those in~\nr{dl1}--\nr{dl7}, is
\be
M=
\left(
\begin{array}{lllllll}
c^4    & s^4    &s2^2/4   &s2^2/4   &s2^2/2  &-c^2 s2  &-s^2 s2 \\ 
s^4    & c^4    &s2^2/4   &s2^2/4   &s2^2/2  &s^2 s2   &c^2 s2 \\ 
s2^2/2 & s2^2/2 &c^4+s^4  &-s2^2/2  &-s2^2   &s2 c2    &-s2 c2 \\ 
s2^2/2 & s2^2/2 &-s2^2/2  &c^4+s^4  &-s2^2   &s2 c2    &-s2 c2 \\ 
s2^2/4 & s2^2/4 &-s2^2/4  &-s2^2/4  &c^4+s^4 &s2 c2/2  &-s2 c2/2 \\ 
s2 c^2 &-s2 s^2 &-s2 c2/2 &-s2 c2/2 &-s2 c2  &c^2-s2^2 &s2^2-s^2 \\ 
s2 s^2 &-s2 c^2 & s2 c2/2 &s2 c2/2  &s2 c2   &s2^2-s^2 &c^2-s2^2 \\ 
\end{array}
\right).
\ee

\section{Integrating out the heavy Higgs doublet}
\la{heavyH}

In eq.~\nr{angle} the angle $\alpha$ has been chosen such
that the field $\phi$ is light at the phase transition point, 
as can be seen from~\nr{d2m1}.
Then the heavy field $\theta$ can be integrated out.
The expansion parameter is
\be
\frac{g_3^2}{4\pi m_\theta}, 
\ee
which is very small in the cases we are studying
(recall that $m_\theta \gsim m_A,T$).
It should be noted that $g_3^2$ arises for the first time
at 2-loop level, whereas at 1-loop level only the 
scalar self-couplings appear.

When $\theta$ is removed, the resulting theory is just
the 3d SU(2)+Higgs theory:
\ba
L_{\rm 3d} & = & \fr14 F^a_{ij}F^a_{ij} 
+ (D_i\phi)^\dagger(D_i\phi)+
m_\phi^2 \phi^{\dagger}\phi+
\lambda_\phi (\phi^\dagger\phi)^2. \la{phi}
\ea
For this theory there are non-perturbative lattice
results available,
so that one need not go any further with
perturbative methods.

Since the interactions in the starting point, eq.~\nr{phithe}, 
involve vertices of the type $\phi\theta^3$ and $\theta\phi^3$, 
there are non-standard graphs needed in the construction
of the effective theory. Numerically these graphs may not be very important
since the relevant 
coupling constants $\lambda_6, \lambda_7$ are not large
and are suppressed by the large mass $m_\theta$ in the results.
Nevertheless, conceptually the way to include
$\lambda_6, \lambda_7$ has to be addressed.
It should be noted that while in eqs.~\nr{lam6}--\nr{lam7}
$\lambda_6, \lambda_7$ are much smaller than
$\lambda_1,\ldots,\lambda_4$ for small mixing parameters, 
in general this is no longer true in the theory of eq.~\nr{phithe} 
due to the redefinitions of fields in Sec.~\ref{diag}. 

Let us start with the wave function normalizations.
The wave function $\phi$ does not get normalized
in the integration, since at 1-loop level
there are no momentum-dependent contributions
to the 2-point correlator $\langle\phi\phi\rangle$ from 
the heavy modes $\theta$. Due to the graph ($\theta\theta$)
in Fig.~\ref{fig:dimred}.a, 
the wave function $A^a_i$ becomes
\be
\Bigl(A^a_iA^b_j\Bigr)^{({\rm new})}= 
\Bigl(A^a_iA^b_j\Bigr) \biggl(
1+\frac{g_3^2}{48\pi m_\theta}
\biggr).
\ee
The gauge coupling is changed to 
\be
g_3^{2({\rm new})}=g_3^2 \biggl(
1-\frac{g_3^2}{48\pi m_\theta}
\biggr).
\ee

For the scalar mass parameter, the diagram 
in Fig.~\ref{fig:2Hdoub}.a gives
\be
m_\phi^{2({\rm new})}=m_\phi^2-
\frac{m_\theta}{4\pi}(2\lambda_3+\lambda_4).
\ee
The scalar coupling constant receives contributions from the
diagrams in Fig.~\ref{fig:2Hdoub}.b to become
\be
\lambda_\phi=\lambda_1-\frac{1}{8\pi m_\theta}
\Bigl(\lambda_3^2+\lambda_3\lambda_4+\fr12 \lambda_4^2+
2\lambda_5^2+12\lambda_6^2-12\lambda_6\lambda_7
\Bigr).
\ee
The coupling $\lambda_2$ would enter only at 2-loop level.
Let us discuss the result for the coupling
constant in some more detail.

\begin{figure}[tb]

\figtopspace

\epsfysize=\figysize
\centerline{\epsffile[50 290 450 740]{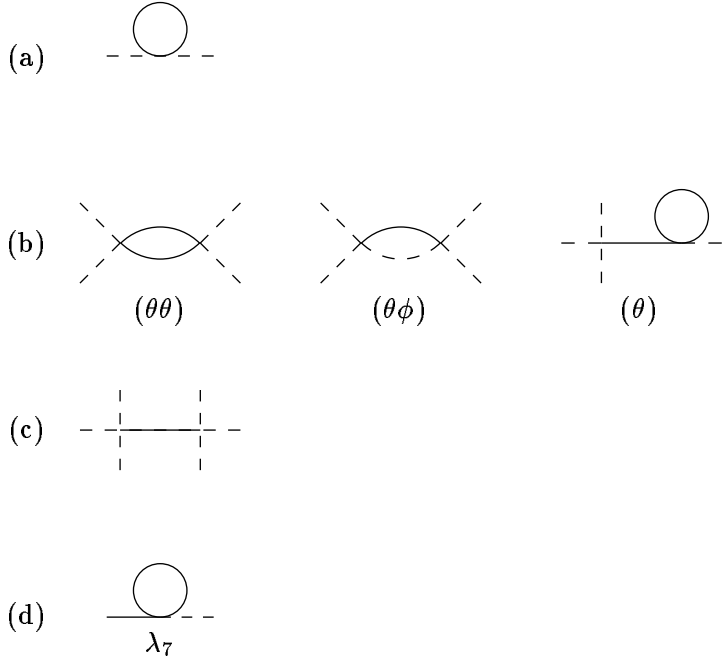}}

\figbottomspace

\caption[a]{
The graphs needed for integrating out the heavy Higgs doublet from 
the 3d two Higgs doublet model. The solid line represents
the heavy field $\theta$. Graph (a) is a contribution
to the mass parameter $m_\phi^2$, graphs (b) 
are contributions to the scalar self-coupling $\lambda_\phi$, 
(c) is an induced 6-point function, and (d) is a mixing
term generated at 1-loop level.}
\la{fig:2Hdoub}
\end{figure}

The contributions involving $\lambda_3, \lambda_4, \lambda_5$ 
come from graphs of the type
($\theta\theta$) in Fig.~\ref{fig:2Hdoub}.b, 
and are standard. 

The contribution proportional 
to $\lambda_6^2$ comes from the graph ($\theta\phi$),
involving a light field in the internal line. 
In principle, 
one might think that $\lambda_6$ contributes
at order $\lambda_6^2$ only to the 6-point function
depicted in Fig.~\ref{fig:2Hdoub}.c. Such a contribution, 
however, has a momentum-dependence:
\be
c_6\sim \frac{\lambda_6^2}{p^2+m_\theta^2}. \la{6po}
\ee 
To construct a local effective theory, one wants
to expand in the momenta. 
Naively one might think that it is justified to expand eq.~\nr{6po}
everywhere in $p^2/m_\theta^2$ since the effective theory 
only involves the mass scale $m_\phi \ll m_\theta$.
This naive procedure is wrong, since if the $\theta$-propagator
is expanded before integration
in the graph $(\theta\phi$) 
of Fig.~\ref{fig:2Hdoub}.b, 
one only gets the suppressed contribution 
\be
\lambda_6^2\frac{m_\phi}{m_\theta^2}.
\ee
In reality, the dominant contribution 
of ($\theta\phi$)
is of order $\lambda_6^2/m_\theta$. Only when this larger
contribution is explicitly included in the reduction step
by the graph ($\theta\phi$), can one expand in the momenta
in eq.~\nr{6po}. Then, in fact, the 6-point operator
$\phi^6$ can be neglected in the effective theory, since
it only leads to contributions suppressed by $m_\phi/m_\theta \ll 1$.

The phenomenon explained is of course the same which appears
in the dimensional reduction step at 2-loop level~\cite{j,klrs1}
when one is comparing a naive integration over 
non-zero Matsubara frequencies 
with a matching procedure for the construction of an effective
theory. The former method, containing exclusively heavy modes in the 
internal lines, leads to a non-local theory.
The latter method leads to a local effective theory, 
but light fields have to be included in the internal lines
of some graphs.
In the present case, the difference appears already at 1-loop level.
In principle, a systematic way to account for these effects is to 
split the light fields into low-momentum modes $\phi_{|p|<\Lambda}$
and high-momentum modes $\phi_{|p|>\Lambda}$; then only
the heavy fields and the high-momentum modes of the light
fields need be included in the internal lines.

The contribution ($\theta$) in Fig.~\ref{fig:2Hdoub}.b is even more
exotic than ($\theta\phi$). 
It cannot even be generated from an effective
potential for the $\phi$-field alone as the other contributions, 
since the graph is reducible.
This contribution arises because the vertex involving
$\lambda_7$ induces a mixing between $\phi$ and $\theta$ at 1-loop 
level, as shown in Fig.~\ref{fig:2Hdoub}.d. This mixing does not 
vanish in the limit that $m_\theta$ is large, but 
grows as $\lambda_7 m_\theta$. 
Nevertheless, it is still possible to construct order by order
an effective theory of the type in~\nr{phi},
containing the light fields only and
giving the same light Green's functions as the original theory.

In the configuration ($\theta$) in 
Fig.~\ref{fig:2Hdoub}.b, the induced 
mixing contributes to the 4-point
function of the $\phi$-fields at the same order of magnitude
as ($\theta\theta$), ($\theta\phi$). To reproduce
this contribution in the theory of eq.~\nr{phi}, the graph ($\theta$)
has to be included in the reduction step. Working at 1-loop
order, one may expand the momentum dependence of this graph, 
but going to 2-loop order, the graph
obtained from ($\theta$) by contracting the rightmost light
field with one of the other light fields has to be included
in the calculation of the mass parameter $m_\phi^2$
to order $\lambda^2$.

Finally, let us recall from Sec.~\ref{sm} that 
the parameter relevant for baryogenesis in the theory
of eq.~\nr{phi} is the dimensionless 
ratio $x=\lambda_\phi/g_3^2$ at the phase transition point.
The temperature dependence of 
$\lambda_\phi/T, g_3^2/T$ is weak: at the dimensional reduction
step the dependence comes only through logarithmic
1-loop corrections. In the heavy scale integrations 
a larger dependence is induced 
since e.g.\ $h_{t3}^2/m_{U3} \sim
h_t^2 T/\sqrt{m_U^2+\gamma T^2}$ depends on $T$.
We estimate the critical temperature from the condition
\be
m_\phi^2=0, \la{mis0}
\ee
which gives sufficient accuracy for the present purpose.
In particular, note 
that imposing eq.~\nr{mis0} in the 3d effective theory
generally involves
the next-to-leading corrections~\cite{a} to $T_c$
in terms of the original coupling constants, 
arising from the heavy 3d modes.
The numerical results obtained for $x$ 
are discussed in Sec.~\ref{numres}.

\section{Vacuum renormalization}
\la{vacren}

To complete the program of dimensional reduction, one has 
to fix the running parameters appearing
in Sec.~\ref{dimred} in terms of zero
temperature pole masses and cross sections. 
We reiterate first the general strategy adopted in the
present paper. 

1. The Higgs mass parameters $m_1^2(\bmu)$, $m_2^2(\bmu)$, 
$m_{12}^2(\bmu)$ are determined in this Section. 

2. The running of $m_Q^2(\bmu), 
m_U^2(\bmu), m_D^2(\bmu)$ can be read from 
eqs.~\nr{mmQ3}--\nr{mmD3}.
Since the squark masses are not known at present, we 
do not fix the running parameters in terms of pole 
masses here.  However, 
once the squark masses have been measured, the 
renormalization must be properly performed, 
since especially the possibly small stop
mass parameter $m_U^2$ has a significant effect on 
the phase transition. In the present work 
we use tree-level values for
$m_Q^2(\bmu_T), 
m_U^2(\bmu_T), m_D^2(\bmu_T)$. 

3. For the present type of investigations, 
the running gauge coupling is most conveniently
fixed in terms of the muon lifetime~\cite{klrs1}. 
In the complete MSSM, the result could be extracted from 
a calculation of the type in~\cite{deltar}.
However, as stated before, 
within the present theory the 
gauge coupling appearing is not 
universal beyond tree-level 
if other than squark and quark loops are included in vacuum 
renormalization. Hence there is no use in going into
elaborate investigations; we will rather
fix  $g(\bmu=200 {\rm GeV})=2/3$
and include only the running due to quarks and squarks. 

4. The U(1) gauge coupling is taken at tree-level, and
is fixed to be $g'=1/3$. At 1-loop level no difference is made
between $g^2$ and $\tilde{g}^2=g^2+g'^2$. Since
vacuum renormalization is related to the neutral sector
of the theory, we will use the numerical value of 
$\tilde{g}^2$ in the loops calculated with
the SU(2) interactions. 

5. The Yukawa couplings $h_t(\bmu), h_b(\bmu)$ can 
in principle be fixed in 
terms of the top and bottom pole masses. 
However, within the approximations
of the present paper, there is no universal $h_t(\bmu)$ at
1-loop level. Hence we fix also $h_t, h_b$
at tree-level. For fixed $\tb=v_2/v_1$ and $\bmu$, we take
\be
h_t = \frac{\tilde{g}}{\sqrt{2}}\frac{m_t}{m_Z}\frac{1}{\sin\beta}, \quad
h_b = \frac{\tilde{g}}{\sqrt{2}}\frac{m_b}{m_Z}\frac{1}{\cos\beta}. 
\ee

6. The mixing parameters are also running parameters. 
However, they are not very important for the phase transition, 
at least if small, and they are not known. 
We fix them too at the tree-level through
\ba
w_s & = & -\mu h_t, \la{cub1}\\
e_s & = & \mu h_b, \la{cub2} \\
u_s & = & -h_t\tilde{A}_t - w_s\cot\beta, \\ 
d_s & = & -h_b\tilde{A}_b + e_s\tan\beta. \la{cub4}  
\ea
It general, $w_s$ and $e_s$ 
are arbitrary soft supersymmetry breaking parameters 
and hence $\mu$ may be interpreted as something different from 
the supersymmetric mass parameter in the superpotential.
With the conventions in~\nr{cub1}--\nr{cub4}, the squark mass matrices
in the broken phase,
\be
{\cal M}^2_{U}=\left(
\begin{array}{ll}
m_{U1}^2 & m_{U12}^2 \\
m_{U12}^2 & m_{U2}^2   
\end{array}
\right)
\la{matrLR}
\ee
and analogously for ${\cal M}^2_D$, are 
given by
\ba
m_{U1}^2 & = & m_{\tilde{t}_L}^2
= m_Q^2+m_t^2+\fr12 m_Z^2 \cos \!2\beta\, ,\quad
m_{U2}^2= m_{\tilde{t}_R}^2 
= m_U^2+ m_t^2, \la{mtL} \\
m_{U12}^2 & = & \frac{1}{\sqrt{2}}(u_s v_2+w_s v_1) =
-m_t\tilde{A}_t, \la{mtLR} \\ 
m_{D1}^2 & = & m_{\tilde{b}_L}^2
= m_Q^2+m_b^2-\fr12 m_Z^2 \cos \!2\beta\, ,\quad
m_{D2}^2= m_{\tilde{b}_R}^2 =
m_D^2+ m_b^2, \la{mbL} \\
m_{D12}^2 & = & \frac{1}{\sqrt{2}}(e_s v_2-d_s v_1) =
m_b\tilde{A}_b. \la{mbLR}
\ea

We next concentrate on fixing 
$m_1^2(\bmu)$, $m_2^2(\bmu)$, 
$m_{12}^2(\bmu)$ in terms of the pole
masses $m_h$, $m_A$, $m_Z$ of the lightest CP-even
Higgs particle $h$, the CP-odd Higgs particle $h_A$ 
and the Z-boson $Z_\mu$, respectively. 
Going to the 
classical broken minimum determined by $m_1^2(\bmu)$, $m_2^2(\bmu)$, 
$m_{12}^2(\bmu)$ and the scalar couplings, 
one can calculate the tree-level
masses $m_h^2(\bmu)$, $m_A^2(\bmu)$, $m_Z^2(\bmu)$:
\ba
m_A^2(\bmu) & = & m_1^2(\bmu)+m_2^2(\bmu),\nn \\ 
m_Z^2(\bmu) & = & -m_A^2(\bmu)+
\bigl[{m_2^2(\bmu)-m_1^2(\bmu)}\bigr]/{\cbb(\bmu)}, \la{mZ2mu} \\
m_h^2(\bmu) & = & \fr12\biggl[
m_A^2(\bmu)+m_Z^2(\bmu)-
\sqrt{\Bigl(m_A^2(\bmu)+m_Z^2(\bmu)
\Bigr)^2-4 m_A^2(\bmu)m_Z^2(\bmu)\cos^2 \!2\beta\,(\bmu)}
\biggr]. \nn
\ea
Here
\be
\sin \!2\beta\,(\bmu)=-\frac{2 m_{12}^2(\bmu)}
{m_1^2(\bmu)+m_2^2(\bmu)}.
\ee
Adding to the tree-level expressions the 1-loop self-energies 
$\Pi_h(k^2,\bmu)=\langle h(k)h(-k)\rangle$, 
$\Pi_A(k^2,\bmu)=\langle h_A(k)h_A(-k)\rangle$,
and the transverse part  
$\Pi_Z(k^2,\bmu)$ of the 
Z-boson self-energy $\langle Z_\mu(k)Z_\nu(-k)\rangle$, 
evaluated at the corresponding poles,
gives the physical $\bmu$-independent masses:
\ba
m_h^2 & = & m_h^2(\bmu) - \Pi_h(-m_h^2,\bmu), \nn \\
m_A^2 & = & m_A^2(\bmu) - \Pi_A(-m_A^2,\bmu), \la{vaceq} \\
m_Z^2 & = & m_Z^2(\bmu) - \Pi_Z(-m_Z^2,\bmu). \nn
\ea
{}From these equations one can solve for 
$m_h^2(\bmu), m_A^2(\bmu), m_Z^2(\bmu)$ for given 
$m_h^2, m_A^2, m_Z^2$. 
Using the expressions inverse to~\nr{mZ2mu}, 
\ba
m_1^2(\bmu) & = & \fr12 \Bigl\{
m_A^2(\bmu) -\bigl[m_A^2(\bmu)+m_Z^2(\bmu)\bigr]\cbb (\bmu)\Bigr\},\nn \\
m_2^2(\bmu) & = & \fr12 \Bigl\{
m_A^2(\bmu) +\bigl[m_A^2(\bmu)+m_Z^2(\bmu)\bigr]\cbb (\bmu)\Bigr\}, 
\la{mmmu} \\
m_{12}^2(\bmu) & = & -\fr12 m_A^2(\bmu) \sbb(\bmu) \nn
\ea
where 
\be
\sin \!2\beta\,(\bmu)=\frac{1}
{m_A(\bmu)m_Z(\bmu)}\sqrt{{[m_A^2(\bmu)-m_h^2(\bmu)]
[m_Z^2(\bmu)-m_h^2(\bmu)]}}, \la{ssb} 
\ee
one then gets the desired expressions 
for $m_1^2(\bmu), m_2^2(\bmu), m_{12}^2(\bmu)$.
Note that at tree-level the running parameters
equal the physical parameters and eq.~\nr{ssb} implies
the known relation $m_h\le {\rm min}(m_A,m_Z)$, but
at 1-loop level the running parameters are
no longer pole masses and consequently
the Higgs pole mass can be 
considerably larger than implied by the tree-level bound.

A few comments are in order. 
First, in solving eqs.~\nr{vaceq}
it is technically convenient to keep $\tb$ fixed 
rather than the Higgs mass. The parameter $\tb$ is not 
a physical quantity, though, and depends on the gauge, 
on the scheme and on $\bmu$. When
$\tb$ is fixed, 
$m_h$ is not an input parameter any more but comes out 
as a result from~\nr{vaceq} 
for given $\tb$ and $\bmu$. 

Second, we choose to 
use the physical pole masses as the mass parameters 
in the self-energies $\Pi(k^2,\bmu)$. This reduces
the higher loop $\bmu$-dependence of the result. 
In this procedure, one also needs the unknown pole mass of the 
heavier CP-even Higgs mass $m_H$ (in the tadpole diagrams). 
The corresponding self-energy 
can be trivially obtained from $\Pi_h(k^2,\bmu)$, 
see Appendix~A. 
Then one has 
to add the additional unknown $m_H$ and the 
additional equation 
\ba
m_H^2 & = & m_H^2(\bmu) - \Pi_H(-m_H^2,\bmu) \la{mHeq}
\ea
to the three equations~\nr{vaceq}. 
This system of four equations 
[fixed: $\bmu$, $\tb$, $m_A$, $m_Z$; 
unknown: $m_h$, $m_H$, $m_1^2(\bmu)$, $m_2^2(\bmu)$;
eqs: \nr{vaceq}, \nr{mHeq}]
is easily solved by iteration. 

Third, for the present purpose it is sufficient
to work strictly at 1-loop level. 
Numerically, it is important in some regions of the
parameter space to include higher-order corrections, arising
for instance from the mixing induced at 1-loop level between 
the tree-level mass eigenstates $h$ and $H$~\cite{cpr}.

The 1-loop self-energies have been calculated
in the literature~\cite{cpr,don} in detail, but
for completeness we also display the formulas used here
in Appendix A. 
Using the tree-level expressions in terms of $m_1^2$, $m_2^2$, $m_{12}^2$, 
$m_Q^2$, $m_U^2$, $m_D^2$ for the parameters appearing,
it is straightforward to
verify explicitly that the $\bmu$-dependences produced for
$m_1^2$, $m_2^2$, $m_{12}^2$ through~\nr{vaceq}, \nr{mmmu}
agree to leading order with the ones in~\nr{mm13}--\nr{mm123}.
It should be noted, however, that numerically the remaining higher order
$\bmu$-dependence may be as important as the leading order one.
Fortunately, the $\bmu$-dependence of the coupling constants determining the
Higgs mass bound is smaller than that of the mass parameters, see below.

\section{Numerical results}
\la{numres}

Combining the results for vacuum renormalization 
with the formulas for dimensional reduction 
and heavy scale integrations one can study the
values of $x$ in the phenomenologically allowed
part of the MSSM parameter space.
The phenomenological constraints on the squark sector, 
relevant for the present analysis, have been discussed
in~\cite{eqz,beqz,cqw}. First, there exist lower bounds
on the masses of the weakly interacting squarks. This gives
a lower bound on $m_Q$, the stronger constraint arising from 
sbottom. More important, the relative mass splitting
\be
(m_{\tilde{t}_L}^2-m_{\tilde{b}_L}^2 )/{m_{\tilde{t}_L}^2}
\ee
of the left-handed squarks is constrained
by the parameter $\Delta\rho(\tilde{t},\tilde{b})$ 
not to be too large. Since 
$m_{\tilde{t}_L}^2$ contains $m_t^2$ and 
$m_{\tilde{b}_L}^2$ contains $m_b^2$ and
both contain $m_Q^2$, this also
acts as a lower limit for $m_Q^2$. 
We will take $m_Q=300$ GeV which should satisfy the
phenomenological constraints for our reference value $m_t=175$ GeV 
within the accuracy of the present calculation.
At the same time the chosen value of $m_Q$ still lies 
within the applicability
of the high-temperature expansion.

For the right-handed stop mass parameter $m_U^2$ there appear to 
be no phenomenological lower bounds apart from the  
absence of charge and colour breaking~\cite{cqw}. 
On the other hand, one cannot take too small values
within the applicability of the integrations in Sec.~\ref{squarks}, 
since then the expansion parameters in eq.~\nr{expU3} grow large. We 
take $m_U=100$ GeV as a reference value. 
The mixing parameters are taken to be zero at the reference point, 
$\tilde{A}_t=\tilde{A}_b=\mu=0$ (we stress again that
$\mu$ in~\nr{cub1}--\nr{cub2} is not really the
supersymmetric mass parameter affecting the chargino
and neutralino masses, and hence a small value for
it is acceptable). 

We fix the renormalization scale used in vacuum renormalization 
and dimensional reduction to be $\bmu=200$ GeV. The dependence of
physical quantities (such as $x$) on $\bmu$
is formally of higher order
than the accuracy of the present calculation.
In practice there is some dependence which may be used 
to estimate the accuracy of the results, see below. 
This dependence arises for instance since $h_{t3}$
depends on $\bmu$, having been fixed at tree-level.

We will next vary the CP-odd Higgs mass
$m_A$ between 50 and 300 GeV and inspect the values
obtained for $x$, for different values of the lightest
CP-even pole Higgs mass~$m_h$. In particular, the dependence
on $m_t, m_U$ and the mixing parameters $\tilde{A}_t, \mu$
around the reference point is of interest. 

\begin{figure}[tb]

\figtopspace

\epsfysize=\figysize
\centerline{\epsffile{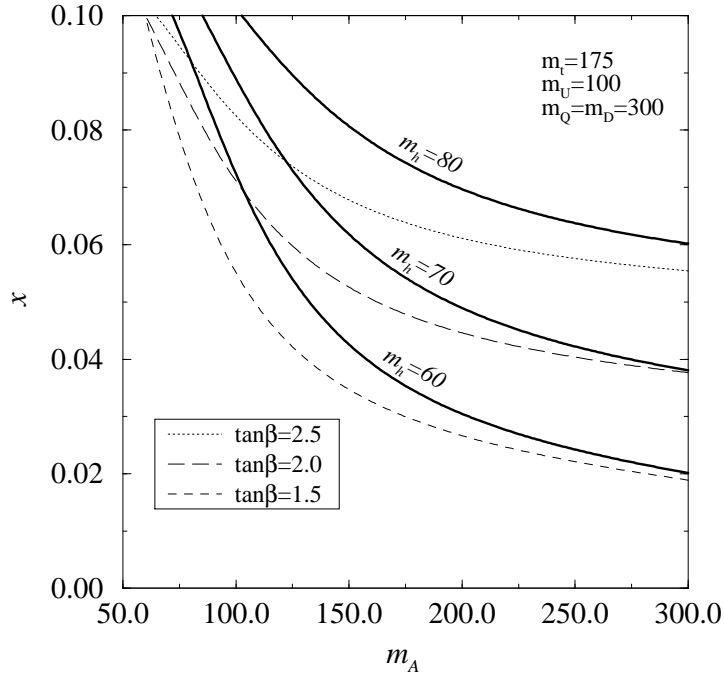}}

\figbottomspace

\caption[a]{
The effect of the CP-even and CP-odd Higgs masses
$m_h$ and $m_A$ on $x$ ($x$ is defined in eq.~\nr{bound}). 
All the numbers are in GeV. 
The mixing parameters have been set to zero. With the thin lines, 
we show as an alternative parametrization
the value of $x$ as a function of 
$\tb$ (at $\bmu=200$ GeV) in the present scheme.}
\la{fig1}
\end{figure}

In Fig.~\ref{fig1} the value of $x$ is shown as a function of $m_A$ 
for the reference set of parameters with three values of $\tb$ 
(thin lines) and three values of the lightest Higgs mass $m_h$
(thick lines). We recall from eq.~\nr{bound}
that the requirement for a strong enough phase
transition to sufficiently suppress the sphaleron rate in the 
broken phase is $x<0.03-0.04$. First, we 
notice that the best region for baryogenesis is a heavy CP-odd
Higgs particle, as is already known~\cite{beqz}. 
Second, for the reference parameters, even
the region $x<0.03$ can be reached with a sufficiently small Higgs
mass (although then our results are less reliable, see
Sec.~\ref{squarks} and below). 
This should be contrasted with the situation in the 
Standard Model where it appears that no Higgs mass is possible~\cite{klrs2}.
Hence the situation has definitely improved in the MSSM.
However, the Higgs mass needed in the MSSM would be rather small, 
$m_h\lsim 70$ GeV. This might soon be excluded
experimentally.

In Fig.~\ref{fig2} the effect of the top mass is shown, 
for fixed $\tb$ (thin lines) and fixed $m_h=70$ GeV 
(thick lines). In contrast to the Standard Model 
(see Fig.~27 in~\cite{klrs2}) a large top mass makes the
situation more favourable for baryogenesis. The reason for 
the difference is that in the MSSM 
the top Yukawa coupling also appears 
in the dimensionally reduced 3d effective
theory through squark interactions. The corrections induced
for the scalar self-coupling are large and negative as 
seen in eqs.~\nr{lam2}--\nr{lam4}. However, since the corrections are 
large, they are also sensitive to the precise value of $h_{t3}$.
A way to estimate the reliability of the results is their $\bmu$-dependence, 
which exists since $h_{t3}$ is fixed only at tree-level. By varying
$\bmu$ from 200 Gev to 300 GeV for fixed $m_h$, the change in 
$x$ is less than 3\% for $m_t\leq175$ GeV. For $m_t=190$ GeV the 
change is about 10\%. Hence the calculation
becomes less reliable for large top mass.

\begin{figure}[t]

\figtopspace

\epsfysize=\figysize
\centerline{\epsffile{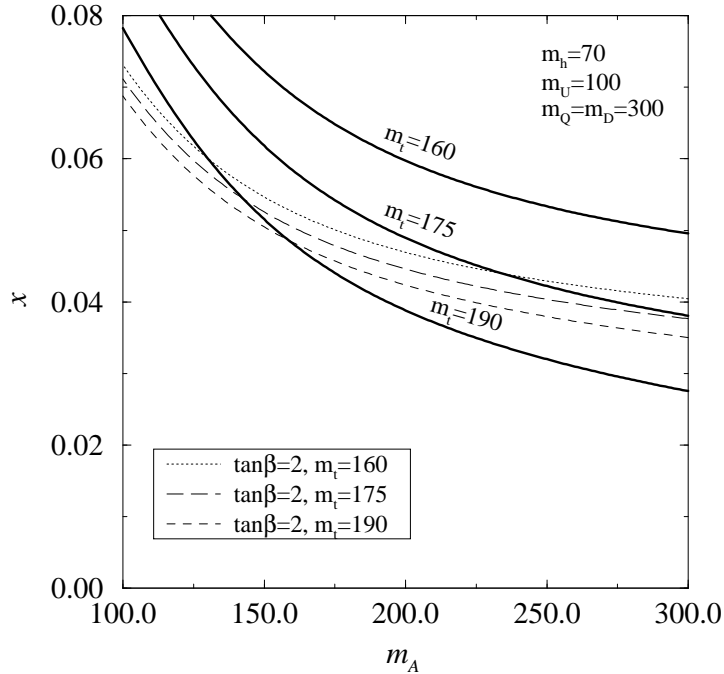}}

\figbottomspace

\caption[a]{
The effect of the top mass on $x$. Here the top
mass is taken at tree-level, in accordance
with the other uncertainties in the calculation. 
The thick lines are for constant $m_h=70$ GeV, the thin lines 
for constant $\tb =2.0$. The mixing parameters
have been set to zero.}
\la{fig2}
\end{figure}

It should be noted that we have kept $m_Q$ fixed when varying $m_t$. 
In fact, if $m_t$ is larger, then also $m_Q$ is likely to be larger, 
in order to keep the mass difference of left-handed stops and
sbottoms small as required by phenomenological 
constraints~\cite{eqz,beqz}. 
This effect would compensate for the 
increase in the strength of the transition with $m_t$~\cite{e}.
In principle, $m_t$ also directly affects  
the running of $m_Q^2(\bmu), m_U^2(\bmu)$ to $\bmu=\bmu_T$, 
but these effects have been neglected here.

In Fig.~\ref{fig3} the 
effect of varying the squark mass parameter $m_U^2$ is shown, 
again separately for fixed $\tb$ and for a fixed Higgs mass. A smaller
$m_U$ makes the situation more favourable, as was already noted 
in~\cite{eqz,beqz,cqw}. In~\cite{cqw} even negative
values for $m_U^2$ were considered. Here we cannot
go to that region since then the squarks are not heavy any more 
and the effective theory is different, see Sec.~\ref{Utheory}. 
Nevertheless, one can see how the effect starts to arise.
It is seen that for $m_U=50$ GeV even a Higgs mass in the region
$m_h\sim 75$ GeV seems possible. 

\begin{figure}[t]

\figtopspace

\epsfysize=\figysize
\centerline{\epsffile{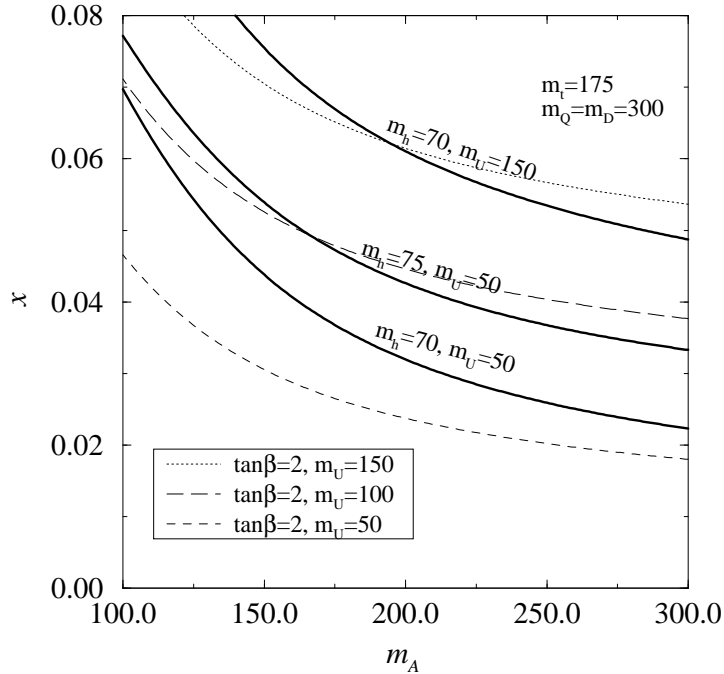}}

\figbottomspace

\caption[a]{
The effect of $m_U$ on $x$ for constant $m_h$ (thick lines)
and $\tb$ (thin lines). Note that at tree-level, 
the right-handed stop mass $m_{\tilde{t}_R}$ 
at zero temperature is given by $m_{\tilde{t}_R}^2 = m_U^2+ m_t^2$
for vanishing $\tilde{A}_t$ and $g'$, 
see~\nr{matrLR}--\nr{mtLR}.}
\la{fig3}
\end{figure}

Finally, in Fig.~\ref{fig4} the effect of the mixing parameters
is presented. Comparing with Fig.~\ref{fig1}, one
can see that $\mu$ has very little effect ($x$ is
just slightly reduced at $m_A\sim 300$ GeV). Indeed, 
for large $m_A$ the mixing is determined exclusively by 
the combination $\tilde{A}_t$ appearing in the 
squark mass matrix. The effect of $\tilde{A}_t$ is that
a large value makes $x$ larger. 
Phenomenologically, this is somewhat unfortunate~\cite{cqw}
since one might wish to have a non-zero
mixing in order to get smaller squark masses, 
which might help with the $R_b$-problem. The sbottom mixing 
parameter $\tilde{A}_b$ has practically no effect at all. 

\begin{figure}[t]

\figtopspace

\epsfysize=\figysize
\centerline{\epsffile{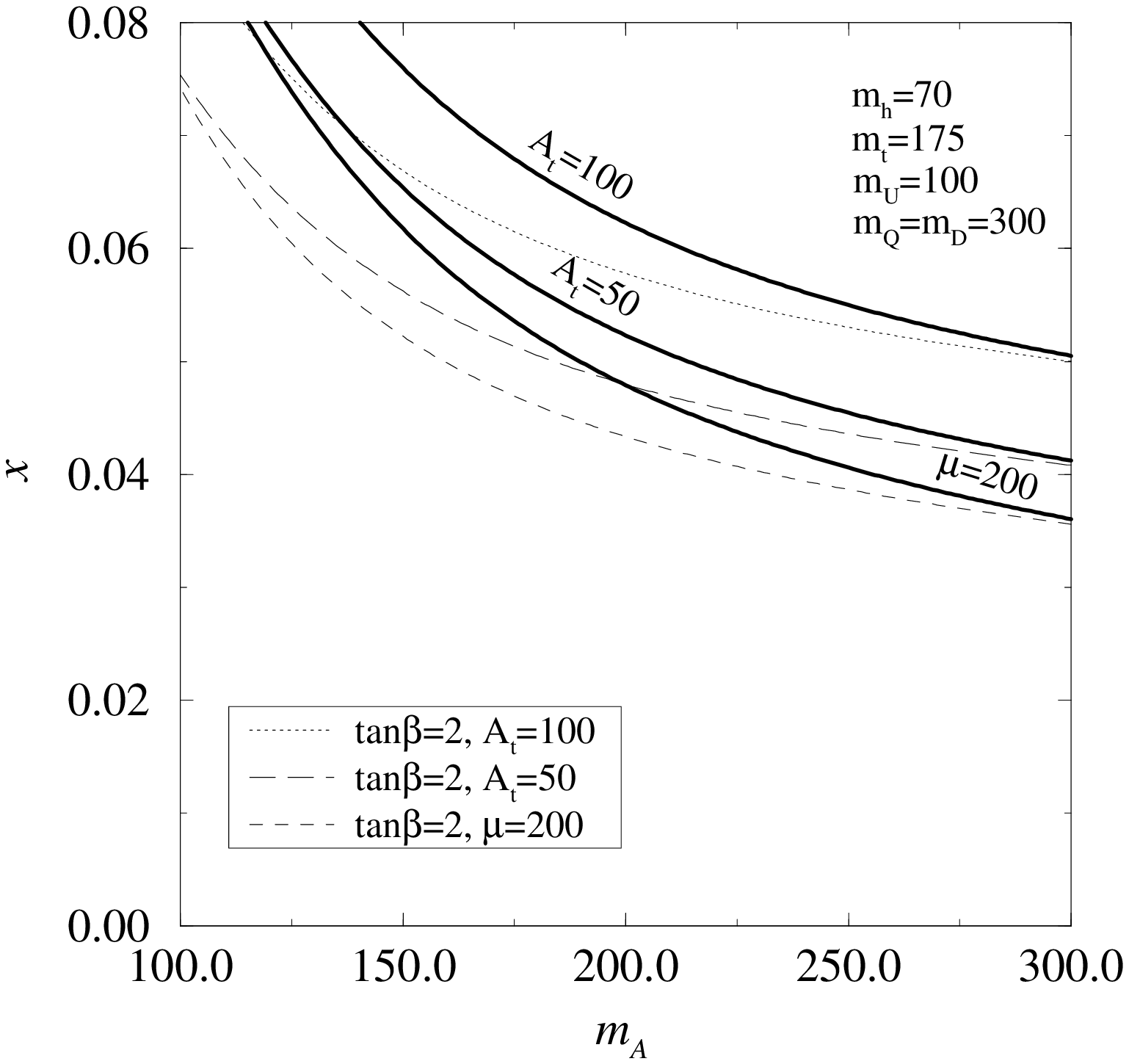}}

\figbottomspace

\caption[a]{
The effect of the mixing parameters on $x$ for constant $m_h$ (thick lines)
and $\tb$ (thin lines). The notation $A_t$ in the figure
stands for $\tilde{A}_t$. If $\mu=0$ or if $m_A$ is large 
so that $\mu$ has little effect, the results are symmetric under
$\tilde{A}_t\to-\tilde{A}_t$ so that only positive
values are shown. For smaller $m_A$, the increase
in $x$ is smallest when the signs of $\mu$ and $\tilde{A}_t$ are the same.}
\la{fig4}
\end{figure}

The overall conclusion is that for $m_h\lsim m_W$
and small mixing, the transition might be strong enough
for $m_U\lsim 50$ GeV and $m_A\gsim 200$ GeV\footnote{
In~\cite{ck} it was proposed that another favourable region
is at small $m_A$, independent of $\tb$. From Fig.~\ref{fig1}
it can be seen that $x$ is indeed almost independent of $\tb$
for $m_A\sim 50$ GeV (this feature persists also for 
values of $\tb$ larger than shown in Fig.~\ref{fig1}). However, $x$
is still much too large and $m_h$ much too small, as can be
seen from the $m_h=60$ GeV curve. Hence the effect proposed
in~\cite{ck} does not take place  close to our reference point.
}. 
If the mixing is larger, one would
need an even smaller $m_U$. These results agree to large extent
with~[4--7]. 
In any case, the situation
has certainly improved with respect to the Standard Model\footnote{
It should also be noted than in the Standard Model there is 
a critical Higgs mass above which the phase transition ceases
to be of first order~\cite{klrs3}. In the MSSM, on the contrary, 
there exists an upper bound on $m_h$, and in some cases (e.g.\
in the vicinity of our reference point) all possible Higgs masses result
in a first order transition.}. 


\section{The effective 3d theory in the case of light right-handed stops}
\la{Utheory}
 
It has been stressed in~\cite{cqw,e} that small 
values of $m_U^2$ are phenomenologically allowed 
and are favourable for electroweak baryogenesis. On the other 
hand, it was pointed out in Sec.~\ref{squarks} that, if $m_U^2$ is 
small, one cannot integrate out the right-handed squarks and
the relevant effective 3d theory is not the simple 
SU(2)+Higgs theory discussed in the previous Sections. 
In this Section we discuss
in more detail at which point the 3d integration 
is no longer reliable and what the relevant 3d theory is then. 
We also indicate some non-perturbative 
effects which might arise beyond the perturbative ones
discussed in~\cite{cqw,e}.

In~\cite{cqw}, the case $m_{U3}\sim 0$ was investigated. 
{}From eq.~\nr{expU3} it is clear that then the integration
does not work at all. To get a quantitative estimate of the
$m_{U3}$ still allowed, one should compare
1-loop and 2-loop contributions to different parameters 
of the effective theory: $h_{t3}$-corrections first arise at 
1-loop level and $g_{S3}$-corrections at 2-loop level, so 
that a comparison of tree-level and 1-loop results does not 
reveal much. The dominant 2-loop contributions were
identified in~\cite{e}.

The dominant 2-loop effect~\cite{e} is due to
graphs of the type 
\be
V_{2l}=-2g_S^2{\cal D}_{\rm SSV}
(m_{\tilde{t}_R},m_{\tilde{t}_R},0) \la{dssv}
\ee
in the notation of~\cite{e,ae}. 
Here essentially
$m^2_{\tilde{t}_R} =  m_{U3}^2+h_t^2 v_2^2/2$ (in~\cite{e}
the coupling constant appearing in this formula is $h_t^2\sin^2\beta$
due to the limit $m_A\to\infty$).
According to eq.~(82) of~\cite{klrs1}, 
the term in~\nr{dssv} affects
the dimensional reduction step of Sec.~\ref{dimred}
only by changing the mass parameter $m_2^2$
by terms of the type
\be
\delta m_2^2 \sim\frac{T^2}{16\pi^2} g_S^2 h_t^2 
\ln\frac{\bmu}{T}.
\ee
These terms are related to the running of $h_t^2(\bmu)$ 
in the 1-loop thermal correction and hence their
inclusion requires 1-loop renormalization of the
top quark mass. In any case, these terms only affect the critical
temperature and thus are not very important.

The (finite) 3d-part of the 2-loop contribution, on the 
other hand, is~\cite{klrs1}
\be
-2 g_S^2 {\cal D}^{\rm 3d}_{\rm SSV}(m_{\tilde{t}_R},m_{\tilde{t}_R},0) =
8 g_S^2 m^2_{\tilde{t}_R} \frac{T^2}{16\pi^2}
\biggl(\ln \frac{\bmu}{2 m_{\tilde{t}_R}}+\fr34 \biggr).
\ee
Now, if $m_{U3}$ is large enough and the transition is not 
too strong, this term can be expanded 
in powers of $h_t^2 v_2^2/(2m_{U3}^2)$. The first term, proportional
to $v_2^2$, changes the mass parameter $m_2^2$ at 2-loop
level, and is not very important. The second term
is quartic in $v_2$ and changes the coupling $\lambda_2$ by
\be
\delta \lambda_2^{2l}=-\frac{g_S^2 h_t^4 T^3}{8\pi^2 m_{U3}^2}.
\la{2lalam2}
\ee
For clarity, we have here kept the coupling constants in 
their 4d normalizations so that powers of $T$ are written
explicitly. The change in eq.~\nr{2lalam2} 
is negative, reducing the coupling constant
$\lambda_2$ and consequently making baryogenesis more
likely. Hence, as long as the expansion converges, this 2-loop
correction works in a favourable direction 
also in the framework of the effective
SU(2)+Higgs theory discussed in Secs.~\ref{squarks}--\ref{heavyH}.
However, when the 
effect becomes stronger, the convergence becomes worse
and the higher order operators generated become important.
The right-handed stops 
can no longer be integrated out but act 
as light degrees of freedom. 

The expansion parameter of $U$-field integration
can be estimated by comparing the 2-loop term 
in eq.~\nr{2lalam2} with the corresponding 1-loop
term in eq.~\nr{alam2}:
\be
\delta \lambda_2^{1l}= -\frac{3}{16\pi}\frac{h_t^4T^2}{m_{U3}}.
\ee
Hence the expansion parameter is  roughly
\be
\frac{\delta\lambda_2^{2l}}{\delta\lambda_2^{1l}}=
\fr23\frac{g_S^2T}{\pi m_{U3}}.
\ee
Consequently, to get convergence one needs $m_{U3}\gsim T$.

What would be the effective theory
if $m_{U3}\lsim T$ and the integration does not converge?
Let us assume that $m_A\to \infty$, 
the squark mixing parameters are small 
and $m_Q$ is relatively large as required
by phenomenological constraints for a 
realistic top mass.
Then all the other squark degrees of freedom 
apart from $U$ can be integrated out in 
the dimensionally reduced 3d theory. What 
remains can be written down immediately using
3d gauge invariance:
\ba
L & = &
\fr14 F^a_{ij}F^a_{ij}+\fr14 G^A_{ij}G^A_{ij} \nn \\
& + & (D_i^w H)^\dagger(D_i^w H)+\tilde{m}_H^2 H^\dagger H+
\lambda_H (H^\dagger H)^2 \nn \\
& + & (D_i^{s~} U)^\dagger(D_i^{s~} U)+\tilde{m}_U^2~ U^\dagger U+
\lambda_U~ (U^\dagger U)^2 \nn \\
& + &  \gamma_3 H^\dagger H U^\dagger U. \la{Uthe}
\ea
Here $D_i^w=\partial_i-i g_3\tau^a A_i^a/2$ and
$D_i^s=\partial_i-i g_{S3}\lambda^A C_i^A/2$
(in this effective theory we have denoted  
the complex conjugate of the original $U$-field by $U$).
At tree-level $\gamma_3=h_t^2 \sin^2\!\beta\, T$, 
$\lambda_U=g_S^2T/6$ and 
$\tilde{m}_U^2=m_{U}^2+(4g_S^2/9+h_t^2/6+h_t^2\ssb/6)T^2$. 

The steps needed for a more precise derivation of the theory
in eq.~\nr{Uthe} are in principle the following. First, 
make dimensional reduction as in Sec.~\ref{dimred}
but in the theory where $m_A\to \infty$. In a precise study, 
it would be important to consistently include all the 1-loop
corrections to the Yukawa couplings $h_t$ appearing
in different places. Second, 
integrate out the heavy fields
$Q, D, A_0, C_0$ as in Sec.~\ref{squarks}. Finally, 
make vacuum renormalization in order to fix
the $\msbar$-parameters in terms of physical parameters.
In particular, one should renormalize 
$m_U^2(\bmu)$ and $h_t(\bmu)$ in addition to 
the Higgs sector parameters by calculating
the stop and top masses at 1-loop level.
All these steps are straightforward and parallel
the ones presented in Secs.~\ref{dimred}--\ref{vacren}. 

Let us stress that perturbatively the theory in 
eq.~\nr{Uthe} reproduces the 1- and 2-loop results 
making the dominant effects in~\cite{cqw,e}. In fact,
the 3d theory also contains a resummation of IR-safe higher-loop
contributions, so that it is expected to be more precise
than direct perturbative calculations in 4d. More important, 
eq.~\nr{Uthe} contains all the IR-problems of the theory
and could be used for 3d Monte Carlo simulations.
No such simulations are available at the moment for the complete
theory. However, one can try to use the knowledge
obtained from simulations of the SU(2)+Higgs sector to 
get some insight into the properties of the complete theory.
We make two guesses.

1. In the simulations of the 3d SU(2)+Higgs theory
it was found that in the symmetric phase the relevant degrees 
of freedom are non-perturbative bound 
states~\cite{klrs2,leip,phtw,dkls}. The mass of e.g.\
the scalar bound state may differ much from the perturbative value, 
let alone from the tree-level value. If $\tilde{m}_U^2$ is positive
at $T_c$ so that the SU(3)-part of eq.~\nr{Uthe} is in 
its symmetric phase, one might expect the same phenomenon
to take place here, only the effects would be
stronger than in the SU(2)-sector. 
This is important since the results of~\cite{cqw,e} strongly depend 
on $\tilde{m}_U^2$ and assume a tree-level value for it.
In particular, the non-perturbative 
mass might be significantly larger than the
perturbative and tree-level masses, in which case
the light degrees of freedom of the theory at the 
phase transition point might again be 
described by the 3d SU(2)+Higgs model.  
This time, however, the derivation of the effective
theory would have to be non-perturbative.

2. The symmetric structure of eq.~\nr{Uthe} opens other
interesting possibilities. At the phase transition, 
$\tilde{m}_H^2$ is close to zero, and if $\tilde{m}_U^2$
is also rather close to zero as proposed in~\cite{cqw},
one might end up in a situation where also the charged and 
coloured field $U$ acquires a non-vanishing expectation 
value at some point during the transition. This kind of 
a multi-stage transition might
naturally alter the mechanism of baryogenesis.
Requiring the {\em absence} of colour 
and charge breaking during the transition, 
some constraints on the parameters were given in~\cite{cqw,ccb}. 

A precise investigation of the possibilities proposed 
will have to wait for a detailed perturbative derivation
and a lattice investigation of the theory in eq.~\nr{Uthe}, 
as well as for experimental data on the values
of the unknown parameters.

\section{Conclusions}
\la{conclu}

We have constructed super-renormalizable
3d effective field theories describing the thermodynamics
of the electroweak phase transition in MSSM. The derivation
of these theories is perturbative and free of IR problems. 
The effective theories can then be used for further perturbative 
investigations if IR problems are believed under control, or
better still, for non-perturbative Monte Carlo studies.  

It was found that in a part of the parameter space,
it is possible to reduce the effective theory to
a 3d SU(2)+Higgs theory for which there already exist
lattice results. However, it generically appears that when
the reduction can be done that far, the transition tends
to get rather weak for realistic Higgs masses. Pushing
the parameters into the region of a stronger transition
(a smaller right-handed stop mass parameter $m_U^2$), the 
convergence of the 3d heavy scale integrations gets worse. 

It hence seems that for a strongly first-order transition, 
the relevant effective 3d theory may be more complicated than 
SU(2)+Higgs. A particularly appealing possibility 
is a model containing an 
SU(2) scalar doublet and an SU(3) scalar triplet. If $m_U^2$
indeed turns out to be small, 
this effective theory should probably be studied
in more detail.
As far as the derivation of the 
theory is concerned, the most important pieces 
missing at the moment are the expressions for
the parameters $h_t(\bmu_T)$ and $m_U^2(\bmu_T')$
in terms of zero-temperature physical parameters
beyond tree-level.
The calculations required are straightforward and
parallel the calculations presented in the present paper. 

\section*{Acknowledgements}

I am grateful to D. B\"odeker, M. Carena, 
K. Kainulainen, 
K. Kajantie, 
A. Patk\'os,
M. Shaposhnikov and C.E.M. Wagner for useful discussions. 
The topic was proposed by K. Kajantie.
This work was partially supported by the University of
Helsinki.

\appendix
\renewcommand{\thesection}{Appendix~~\Alph{section}}
\renewcommand{\theequation}{\Alph{section}.\arabic{equation}}

\section{}

In this appendix we give the formulas used
for vacuum renormalization in Sec.~\ref{vacren}.
More complete expressions
can be found e.g.\ in~\cite{cpr,don} (for 
compact approximation schemes and some 2-loop
corrections, see e.g.~\cite{ep} and references therein).

The calculation of the 1-loop self-energies is organized
as follows. We first shift the fields to the classical
broken minimum. Then the mass eigenstates are identified 
and the 1-loop graphs needed
for $\langle h(k)h(-k)\rangle$, 
$\langle h_A(k)h_A(-k)\rangle$ and 
$\langle Z_\mu(k)Z_\nu(-k)\rangle$ are calculated. 
In particular, the tadpole graphs have to be included
since we are not at the exact quantum minimum. According to 
the general strategy of this paper, only 
quarks and squarks of the third generation
are included in the loops.

For fixed $\tb$, 
the location of the classical broken 
minimum of eq.~\nr{broken} is obtained from
\be
v_1=\frac{2m_Z\cos\beta}{\tilde{g}},\quad
v_2=\frac{2m_Z\sin\beta}{\tilde{g}}.
\ee
At the broken minimum, the mass eigenstates
corresponding to the physical neutral Higgs fields $h, H, h_A$ 
are obtained from the fields in eq.~\nr{h1h2}
with the rotations
\ba
h^1_0 & = & \cos\!\alpha\, h + \sin\!\alpha\, H, \la{h10} \\
h^2_0 & = & -\sin\!\alpha\, h+ \cos\!\alpha\, H, \la{h20} \\
h^1_3 & = & \cos\!\beta\, h_{G_0}-\sin\!\beta\, h_A, \\
h^2_3 & = & \sin\!\beta\, h_{G_0}+\cos\!\beta\, h_A. \la{h23}
\ea
At tree-level
the angle $\alpha$ here is given by
\be
\saa = -\sin \!2\beta\,\frac{m_A^2+m_Z^2}{m_H^2-m_h^2},\quad
\caa = \cos \!2\beta\,\frac{m_A^2-m_Z^2}{m_H^2-m_h^2}.
\ee
At 1-loop level we use the physical pole masses for
$m_A^2$, $m_Z^2$, $m_h^2$, $m_H^2$ and the angle $\alpha$
is determined from the expression for $\caa$.
After the redefinitions~\nr{h10}--\nr{h23}, 
the graphs contributing to $\langle hh\rangle$, 
$\langle h_Ah_A\rangle$ and 
$\langle Z_\mu Z_\nu\rangle$ can easily be identified.

The formulas arising are slightly complicated by the fact that the
left- and right-handed squarks mix. For the mass 
eigenstates, we will use the notation $m_{U\pm}^2$, 
$m_{D\pm}^2$ defined by
\be
m_{U\pm}^2=\fr12\Bigl[
m_{U1}^2+m_{U2}^2\pm\sqrt{(m_{U1}^2-m_{U2}^2)^2+4m_{U12}^4}
\Bigr]
\ee
and correspondingly for $m_{D\pm}^2$,
where $m_{U1}^2,\ldots$ are in~\nr{mtL}--\nr{mbLR}.
We also denote
\be
\delta_U = \frac{1}{m_{U+}^2-m_{U-}^2},
\quad 
\delta_D = \frac{1}{m_{D+}^2-m_{D-}^2}.
\ee

Some standard integrals often appearing are denoted as follows.
For $|m_1-m_2|<k<m_1+m_2$,  
\ba
F_H(k;m_1,m_2) & \equiv & 1-\frac{m_1^2-m_2^2}{k^2}\ln\frac{m_1}{m_2}+
\frac{m_1^2+m_2^2}{m_1^2-m_2^2}\ln\frac{m_1}{m_2} \label{Fkm} \\
& - & \frac{2}{k^2}\sqrt{(m_1+m_2)^2-
k^2}\sqrt{k^2-(m_1-m_2)^2}\arctan\frac{\sqrt{k^2-(m_1-m_2)^2}}
{\sqrt{(m_1+m_2)^2-k^2}}.\nonumber
\ea    
Especially, 
\be
F_H(m_1;m_2,m_2) = 2-2 \sqrt{4 r^2-1}\arctan\frac{1}{\sqrt{4 r^2-
1}},
\ee 
where $r=m_2/m_1$ and $r>1/2$. Outside the displayed
kinematic region, an analytic continuation is needed, 
and from that we only use the real part in calculating
the masses. The imaginary parts arising are small.  
We also define a function $F_Z$ arising in the 
calculation of the Z-boson self-energy:
\ba
F_Z(m_1,m_2) \!\! & = &  
\frac{1}{12 m_Z^2}\biggl\{
2m_1^2(m_1^2-m_2^2+m_Z^2)\ln\frac{m_1}{m_Z}+
2m_2^2(m_2^2-m_1^2+m_Z^2)\ln\frac{m_2}{m_Z} \nn \\
& + & 
\Bigl[m_Z^4-2 m_Z^2(m_1^2+m_2^2)+(m_1^2-m_2^2)^2\Bigr]
\Bigl[1+\ln\frac{m_Z^2}{m_1m_2}-
\frac{m_1^2+m_2^2}{m_1^2-m_2^2}\ln\frac{m_1}{m_2} \nn \\
& + & 
F_H(m_Z;m_1,m_2) 
\Bigr]+\fr23 m_Z^4-3 m_Z^2(m_1^2+m_2^2)-
(m_1^2-m_2^2)^2
\biggr\}.
\ea

The irreducible graphs needed are 
shown in Fig.~\ref{fig:vacren} 
and the tadpole graphs in Fig.~\ref{fig:tadpol}.
We leave out the common factor $1/(16\pi^2)$ in
the formulas below.

\begin{figure}[tb]

\vspace*{-0.5cm}

\epsfysize=12cm
\centerline{\epsffile[50 370 450 720]{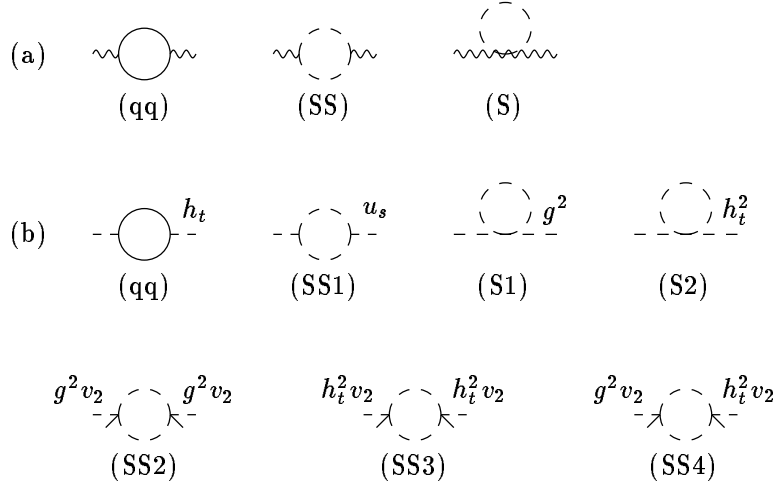}}

\figbottomspace

\caption[a]{
The graphs contributing to (a) the pole
mass $m_Z$ of the Z-boson and (b) the 
masses $m_h, m_H, m_A$ of the CP-even Higgs bosons  
and the CP-odd Higgs boson. In the internal lines, 
solid lines are quarks and dashed lines are squarks of the 
third generation. Examples of couplings appearing
are also shown.}
\la{fig:vacren}
\end{figure}

The contributions 
to $\left.\langle h(k)h(-k)\rangle\right|_{k^2=-m_h^2}$ from 
the graphs in Fig.~\ref{fig:vacren}.b are
\ba
({\rm qq}): \!\! & & \!\!
-3 h_t^2 \ssa
\Bigl[(6m_t^2-m_h^2)\ln\frac{\bmu^2}{m_t^2}+2m_t^2+
(4 m_t^2-m_h^2) F_H(m_h;m_t,m_t)
\Bigr] \nn \\
& & \!\! + \Bigl\{
h_t \to h_b, m_t\to m_b, \ssa\to\cca
\Bigr\}, \la{hqq} \\
({\rm SS1}): \!\! & & \!\!
3(u_s^2\ssa+w_s^2\cca-u_sw_s\saa)\biggl\{ 
\ln\frac{\bmu^2}{m_{U-}m_{U+}} \nn \\
& & \!\!+2 \delta_U^2 m_{U12}^4\Bigl[
F_H(m_h;m_{U-},m_{U-})+
F_H(m_h;m_{U+},m_{U+})\Bigr] \nn \\
& & \!\!+ \delta_U^2 (m_{U1}^2-m_{U2}^2)^2
\Bigl[
1+\delta_U (m_{U1}^2+m_{U2}^2) \ln\frac{m_{U-}}{m_{U+}}+
F_H(m_h;m_{U-},m_{U+})
\Bigr] \biggr\} \nn \\
& & \!\!+ \Bigl\{
u_s\to d_s, w_s\to e_s, \ssa\leftrightarrow\cca,
\saa\to -\saa, U\to D
\Bigr\}, \\
({\rm SS2}): \!\! & & \!\!
\frac{3}{16}\tilde{g}^4(v_2^2\ssa+v_1^2\cca+v_1v_2 \saa)\lmu \nn \\
& & \!\!+\frac{3}{16}\tilde{g}^4(v_2^2\ssa+v_1^2\cca+v_1v_2 \saa)
\delta_U^2 \biggl\{ 
(m_{U2}^2-m_{U-}^2)^2\Bigl[
\ln\frac{m_Z^2}{m_{U-}^2} \nn \\ 
& & \!\!+F_H(m_h;m_{U-},m_{U-})
\Bigr]+ (m_{U2}^2-m_{U+}^2)^2\Bigl[
\ln\frac{m_Z^2}{m_{U+}^2}+F_H(m_h;m_{U+},m_{U+})
\Bigr] \nn \\
& & \!\!+2 m_{U12}^4\Bigl[
1+\ln\frac{m_Z^2}{m_{U-}m_{U+}}+
\delta_U(m_{U1}^2+m_{U2}^2)\ln\frac{m_{U-}}{m_{U+}}+
F_H(m_h;m_{U-},m_{U+})
\Bigr] \biggr\} \nn \\
& & \!\!+\Bigl\{
U\to D
\Bigr\}, \\
({\rm SS3}): \!\! & & \!\!
6 h_t^4 v_2^2\ssa \biggl\{
\ln\frac{\bmu^2}{m_{U-}m_{U+}}+
\fr12\Bigl[
F_H(m_h;m_{U-},m_{U-})+
F_H(m_h;m_{U+},m_{U+})\Bigr]
\biggr\} \nn \\
& & \!\!+\Bigl\{
h_t\to h_b, v_2^2\ssa\to v_1^2\cca, U\to D
\Bigr\}, \\
({\rm SS4}): \!\! & & \!\!
-\fr32 \tilde{g}^2 h_t^2 \Bigl(v_2^2\ssa+\fr12 v_1v_2\saa\Bigr)\lmu \nn \\
& & \!\!-\fr32 \tilde{g}^2 
h_t^2 \Bigl(v_2^2\ssa+\fr12 v_1v_2\saa\Bigr)\delta_U^2\biggl\{  
\Bigl[(m_{U2}^2-m_{U-}^2)^2+m_{U12}^4 \Bigr]
\Bigl[ \ln\frac{m_Z^2}{m_{U-}^2} \nn \\
& & \!\!
+F_H(m_h;m_{U-},m_{U-})\Bigr]+
\Bigl[(m_{U2}^2-m_{U+}^2)^2+m_{U12}^4 \Bigr]
\Bigl[ \ln\frac{m_Z^2}{m_{U+}^2}\nn \\
& & \!\!+
F_H(m_h;m_{U+},m_{U+})\Bigr] \biggr\} +\Bigl\{
h_t\to h_b, v_2^2\ssa \to v_1^2 \cca, U\to D 
\Bigr\},  \\
({\rm S1}): \!\! & & \!\!
\fr34 \tilde{g}^2 \caa m_{U1}^2
\biggl(\lmu +1\biggr) \nn \\
& & \!\!-\fr32 \tilde{g}^2 \caa \delta_U \Bigl[
(m_{U2}^2-m_{U-}^2)m_{U-}^2\ln\frac{m_{U-}}{m_Z}-
(m_{U2}^2-m_{U+}^2)m_{U+}^2\ln\frac{m_{U+}}{m_Z}
\Bigr] \nn \\
& & \!\!-\Bigl\{ U\to D \Bigr\}, \\
({\rm S2}): \!\! & & \!\!
3 h_t^2\ssa (m_{U1}^2+m_{U2}^2)
\biggl(\lmu +1\biggr) \nn \\
& & \!\!-6 h_t^2\ssa\Bigl(
m_{U-}^2\ln\frac{m_{U-}}{m_Z}+
m_{U+}^2\ln\frac{m_{U+}}{m_Z} 
\Bigr) \nn \\
& & \!\!
+\Bigl\{
h_t\to h_b, \ssa\to \cca, U\to D
\Bigr\}. 
\ea
The contributions to 
$\left.\langle h_A(k)h_A(-k)\rangle\right|_{k^2=-m_A^2}$ from 
the graphs in Fig.~\ref{fig:vacren}.b are
\ba
({\rm qq}): \!\! & & \!\!
-3 h_t^2 \ccb
\Bigl[(2m_t^2-m_A^2)\ln\frac{\bmu^2}{m_t^2}+2m_t^2-
m_A^2 F_H(m_A;m_t,m_t)
\Bigr] \nn \\
& & \!\! + \Bigl\{
h_t \to h_b, m_t\to m_b, \ccb\to\ssb
\Bigr\}, \\
({\rm SS1}):  \!\! & & \!\!
3(u_s^2\ccb+w_s^2\ssb-u_sw_s\sbb)\Bigl[ \ln\frac{\bmu^2}{m_{U-}m_{U+}}\nn \\
& & \!\!
+1+\delta_U (m_{U1}^2+m_{U2}^2) \ln\frac{m_{U-}}{m_{U+}}+
F_H(m_A;m_{U-},m_{U+})
\Bigr] \nn \\
& & \!\!+ \Bigl\{
u_s\to d_s, w_s\to e_s, \ccb\leftrightarrow\ssb, 
\sbb\to -\sbb, U\to D
\Bigr\}, \\
({\rm S1}): \!\! & & \!\!
-\fr34 \tilde{g}^2 \cbb 
m_{U1}^2 \biggl(\lmu +1\biggr) \nn \\
& & \!\!+\fr32 \tilde{g}^2 \cbb\delta_U \Bigl[
(m_{U2}^2-m_{U-}^2)m_{U-}^2\ln\frac{m_{U-}}{m_Z}-
(m_{U2}^2-m_{U+}^2)m_{U+}^2\ln\frac{m_{U+}}{m_Z}
\Bigr] \nn \\
& & \!\!-\Bigl\{ U\to D \Bigr\}, \\
({\rm S2}): \!\! & & \!\!
3 h_t^2\ccb (m_{U1}^2+m_{U2}^2)\biggl(\lmu +1\biggr) \nn \\
& & \!\!-6 h_t^2\ccb\Bigl(
m_{U-}^2\ln\frac{m_{U-}}{m_Z}+
m_{U+}^2\ln\frac{m_{U+}}{m_Z} 
\Bigr) \nn \\
& & \!\!
+\Bigl\{
h_t\to h_b, \ccb\to \ssb, U\to D
\Bigr\}. 
\ea
The contributions to the transverse part of 
$\left.\langle Z_{\mu}(k)Z_{\nu}(-k)\rangle\right|_{k^2=-m_Z^2}$ from 
the graphs in Fig.~\ref{fig:vacren}.a are
\ba
({\rm qq}): \!\! & & \!\!
-\fr12 \tilde{g}^2\Bigl[
(3m_t^2-m_Z^2)\ln\frac{\bmu^2}{m_t^2}+\fr13 m_Z^2+
(m_t^2-m_Z^2)F_H(m_Z;m_t,m_t)
\Bigr] \nn \\
& & \!\!+\Bigl\{
m_t\to m_b
\Bigr\}, \\
({\rm SS}): \!\! & & \!\!
-\fr12 \tilde{g}^2\Bigl(3m_{U1}^2-\fr12 m_Z^2\Bigr)\lmu \nn \\
& & \!\!+3\tilde{g}^2\delta_U^2
\Bigl[
(m_{U2}^2-m_{U-}^2)^2F_Z(m_{U-},m_{U-})+
(m_{U2}^2-m_{U+}^2)^2F_Z(m_{U+},m_{U+}) \nn \\
& & \!\!+
2 m_{U12}^4 F_Z(m_{U-},m_{U+})
\Bigr] +\Bigl\{
U\to D
\Bigr\}, \\
({\rm S}): \!\! & & \!\!
\fr32 \tilde{g}^2 m_{U1}^2  \biggl(\lmu +1\biggr) \nn \\
& & \!\!-3\tilde{g}^2\delta_U
\Bigl[
(m_{U2}^2-m_{U-}^2)m_{U-}^2 \ln\frac{m_{U-}}{m_Z}-
(m_{U2}^2-m_{U+}^2)m_{U+}^2 \ln\frac{m_{U+}}{m_Z}
\Bigr] \nn \\
& & \!\!+\Bigl\{
U\to D
\Bigr\}.\la{ZS}
\ea

To the contributions in eqs.~\nr{hqq}--\nr{ZS} one
has to add the tadpole contributions from 
Fig.~\ref{fig:tadpol}, 
since we are working
around the classical minimum.
The tadpole contributions to $\Pi_{h}(k^2,\bmu), 
\Pi_{A}(k^2,\bmu), \Pi_{Z}(k^2,\bmu)$ are
\ba
\Pi_h^{({\rm tad})} & = & -3 \frac{\caa}{\cos \!2\beta\,}\Pi_A^{({\rm tad})}
\la{tadhh} \\
& + & 
\frac{1}{2}\tilde{g}^2 \biggl[
\biggl(
\frac{v_1\ssa+v_2\saa/2}{m_H^2}
\biggr)v_1C_h + \biggl(
\frac{v_2\cca+v_1\saa/2}{m_H^2}
\biggr)v_2S_h
\biggr], \nn \\
\Pi_A^{({\rm tad})} & = & 
-\frac{1}{4}\tilde{g}^2\cos \!2\beta\, \biggl[
\biggl(
\frac{v_1\cca+v_2\saa/2}{m_h^2}+
\frac{v_1\ssa-v_2\saa/2}{m_H^2}
\biggr)v_1C_h \nn \\
& - & \biggl(
\frac{v_2\ssa+v_1\saa/2}{m_h^2}+
\frac{v_2\cca-v_1\saa/2}{m_H^2}
\biggr)v_2S_h
\biggr], \la{tadAA} \\
\Pi_Z^{({\rm tad})} & = & 
\frac{1}{2}\tilde{g}^2\biggl[
\biggl(
\frac{v_1\cca-v_2\saa/2}{m_h^2}+
\frac{v_1\ssa+v_2\saa/2}{m_H^2}
\biggr)v_1C_h \nn \\
& + &  \biggl(
\frac{v_2\ssa-v_1\saa/2}{m_h^2}+
\frac{v_2\cca+v_1\saa/2}{m_H^2}
\biggr)v_2S_h
\biggr]. \la{tadZZ}
\ea
Here 
\ba
S_h & = &
3\Bigl[
2 h_t^2 m_t^2 -u_s^2-e_s^2+
(d_se_s-u_sw_s)\cot\!\beta\, \nn \\
& - & h_t^2({m_{U1}}^2+{m_{U2}}^2)+
\frac{1}{4}\tilde{g}^2({m_{U1}}^2-{m_{D1}}^2)\Bigr]
\biggl(\lmu +1\biggr) \nn \\
& - & 12 h_t^2 m_t^2\ln\frac{m_t}{m_Z}+
6 h_t^2\Bigl(
m_{U+}^2\ln\frac{m_{U+}}{m_Z}+
m_{U-}^2\ln\frac{m_{U-}}{m_Z}
\Bigr) \nn \\
& + & 6 \delta_U u_s(u_s+w_s\cot\!\beta\,)
\Bigl(
m_{U+}^2\ln\frac{m_{U+}}{m_Z}-
m_{U-}^2\ln\frac{m_{U-}}{m_Z}
\Bigr) \nn \\
& + & 
6 \delta_D e_s(e_s-d_s\cot\!\beta\,)
\Bigl(
m_{D+}^2\ln\frac{m_{D+}}{m_Z}-
m_{D-}^2\ln\frac{m_{D-}}{m_Z}
\Bigr) \nn \\
& -  & 
\fr32\tilde{g}^2\delta_U
\Bigl[
(m_{U2}^2-m_{U-}^2)m_{U-}^2 \ln\frac{m_{U-}}{m_Z}-
(m_{U2}^2-m_{U+}^2)m_{U+}^2 \ln\frac{m_{U+}}{m_Z}
\Bigr]  \nn \\
& +  & 
\fr32\tilde{g}^2\delta_D
\Bigl[
(m_{D2}^2-m_{D-}^2)m_{D-}^2 \ln\frac{m_{D-}}{m_Z}-
(m_{D2}^2-m_{D+}^2)m_{D+}^2 \ln\frac{m_{D+}}{m_Z}
\Bigr], \\
C_h & = &  \Bigl\{
h_t\to h_b, m_t\to m_b, u_s\leftrightarrow d_s, 
w_s \leftrightarrow -e_s, \cot\!\beta\,\to\tan\!\beta\,, 
U\leftrightarrow D\Bigr\}. 
\ea

\begin{figure}[tb]

\vspace*{0.0cm}

\epsfysize=6.2cm
\centerline{\epsffile[150 540 420 710]{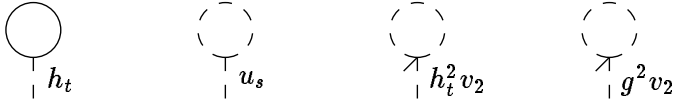}}

\figbottomspace

\caption[a]{
The tadpole graphs needed in vacuum 
renormalization. The closed loops contain quarks and 
squarks of the third generation, and the single line
may contain either of the CP-even Higgs particles.
Representative coupling constants are shown.}
\la{fig:tadpol}
\end{figure}

The contributions to the heavier
CP-even Higgs mass are obtained from the contributions
to the lighter CP-even Higgs mass 
by changing $m_h^2 \leftrightarrow m_H^2$, $\saa \to -\saa$, 
$\caa \to -\caa$, $\ssa \leftrightarrow \cca$
inside the 1-loop formulas for $\Pi_h(-m_h^2,\bmu)$, 
as can be seen from eqs.~\nr{h10}, \nr{h20}.

\section*{Erratum}

\vspace*{-1cm}

\begin{flushright}
(February, 1999)
\end{flushright}

\vspace*{0.2cm}

In Appendix A, some loop contributions were erroneously omitted
from the correlator $\left.\langle h(k) h(-k)\rangle\right|_{k^2=-m_h^2}$, 
Eqs.~(A.12)-(A.18). These contributions, which do not have any 
scale dependence, are proportional to squark mixing parameters and 
are thus very small for the small mixings considered. 
Nevertheless, they should in principle be included.

In the notation of Fig.~7.b, the loops omitted are of the form 



\begin{center}
\begin{picture}(175,40)(0,0)

\DashLine(15,20)(25,20){5}
\DashLine(45,20)(55,20){5}
\DashCArc(35,20)(10,0,180){5}
\DashCArc(35,20)(10,180,360){5}
\Line(45,20)(50,15)

\DashLine(120,20)(130,20){5}
\DashLine(150,20)(160,20){5}
\DashCArc(140,20)(10,0,180){5}
\DashCArc(140,20)(10,180,360){5}
\Line(150,20)(155,15)

\Text(35,0)[]{(SS5)}
\Text(140,0)[]{(SS6)}
 
\Text(20,30)[r]{$u_s$}
\Text(50,30)[l]{$g^2 v_2$}
\Text(125,30)[r]{$u_s$}
\Text(155,30)[l]{$h_t^2 v_2$}

\end{picture}
\end{center}


\noindent
The results for these contributions,
to be inserted between Eqs.(A.16),(A.17), are:
\begin{eqnarray*}
({\rm SS5}): \!\! & & \!\!
-\frac{3}{\sqrt{2}} 
\tilde{g}^2 
(-u_s \sin\!\alpha+w_s \cos\!\alpha)
(v_1 \cos\!\alpha + v_2 \sin\!\alpha) 
\delta_U m_{U12}^2 \biggl\{ \ln\frac{m_{U+}}{m_{U-}}\nn \\
& &  \!\!  + \delta_U 
\Bigl[(m_{U2}^2-m_{U1}^2)
\Bigl(\delta_U (m_{U1}^2+m_{U2}^2) \ln\frac{m_{U+}}{m_{U-}}-1 \Bigr)\nn \\
& & \!\! 
+(m_{U2}^2-m_{U-}^2)F_H(m_h;m_{U-},m_{U-})
+(m_{U2}^2-m_{U+}^2)F_H(m_h;m_{U+},m_{U+})\nn \\
& & \!\!
+(m_{U1}^2-m_{U2}^2)F_H(m_h;m_{U-},m_{U+})\Bigr]\biggr\} \nn \\
& & \!\!
+\Bigl\{
-u_s \sin\!\alpha+w_s \cos\!\alpha \to
d_s \cos\!\alpha + e_s \sin\!\alpha, U\to D 
\Bigr\}, \hspace*{28.5mm} \mbox{(A.16.a)} \\
({\rm SS6}): \!\! & & \!\!
12 h_t m_t  
(-u_s\ssa+\fr12 w_s\saa) \delta_U m_{U12}^2 \nn \\
& & \!\! \times\Bigl[
2 \ln\frac{m_{U+}}{m_{U-}}+
F_H(m_h;m_{U-},m_{U-})-
F_H(m_h;m_{U+},m_{U+})\Bigr] \nn \\
& & \!\!+ \Bigl\{h_t\to h_b, m_t\to m_b, 
u_s \ssa \to -d_s \cca , w_s\to e_s, 
U\to D
\Bigr\}.  \hspace*{9mm} \mbox{(A.16.b)}
\end{eqnarray*}
Numerically, these graphs only affect Fig.~6 where 
non-vanishing mixing was considered. The curves for $\mu=200$ GeV
do not change, since the omitted contributions are proportional 
to $\tilde A_t, \tilde A_b$. The largest effect is for 
$\tilde A_t=100$ GeV: even then, the thin line 
(fixed $\tb$) remains essentially unchanged. The thick line
(fixed $m_h$), comes down by $\sim 0.005$. 
All the conclusions remain unchanged. 

I thank M. Losada for bringing these omissions into my attention.

\end{document}